\documentclass[12pt]{article}

\usepackage{epsf}

\newcommand{\bmat}{\left(\begin{array}}
\newcommand{\emat}{\end{array}\right)}
\def\NPB{Nucl. Phys. B}
\def\PLB{Phys. Lett. B}
\def\PLB{Phys. Lett. B}

\def\yzero{\smash{\hbox{$y\kern-4pt\raise1pt\hbox{${}^\circ$}$}}}

\def\a{\alpha}
\def\b{\beta}
\def\l{\label}

\def\beq{\begin{equation}}
\def\eeq{\end{equation}}
\def\beqa{\begin{eqnarray}}
\def\eeqa{\end{eqnarray}}
\def\t{\times}

\def\-{\hphantom{-}}
\def\ov{\overline}
\def\s2{\frac{1}{\sqrt2}}

\def\beq{\begin{equation}}
\def\eeq{\end{equation}}
\def\beqa{\begin{eqnarray}}
\def\eeqa{\end{eqnarray}}

\def\IF{\relax{\rm I\kern-.18em F}}
\def\II{\relax{\rm I\kern-.18em I}}
\def\IP{\relax{\rm I\kern-.18em P}}
\def\IC{\relax\hbox{\kern.25em$\inbar\kern-.3em{\rm C}$}}
\def\IR{\relax{\rm I\kern-.18em R}}

\def\cp{{\cal P}}

\def\Dsl{\,\raise.15ex\hbox{/}\mkern-13.5mu D} 
\def\IZ{Z\kern-.4em  Z}

 \def\cp#1{\relax\ifmmode {\IP\kern-2pt{}_{#1}}\else $\IP\kern-2pt{}_{#1}$\=fi}

\catcode`\@=11
\newdimen\@rotdimen
\newbox\@rotbox

\def\@vspec#1{\special{ps:#1}}
\def\@rotstart#1{\@vspec{gsave currentpoint currentpoint translate
   #1 neg exch neg exch translate}}
\def\@rotfinish{\@vspec{currentpoint grestore moveto}}
\def\@rotr#1{\@rotdimen=\ht#1\advance\@rotdimen by\dp#1%
   \hbox to\@rotdimen{\hskip\ht#1\vbox to\wd#1{\@rotstart{90 rotate}%
   \box#1\vss}\hss}\@rotfinish}
\def\@rotl#1{\@rotdimen=\ht#1\advance\@rotdimen by\dp#1%
   \hbox to\@rotdimen{\vbox to\wd#1{\vskip\wd#1\@rotstart{270 rotate}%
   \box#1\vss}\hss}\@rotfinish}%
\def\@rotu#1{\@rotdimen=\ht#1\advance\@rotdimen by\dp#1%
   \hbox to\wd#1{\hskip\wd#1\vbox to\@rotdimen{\vskip\@rotdimen
   \@rotstart{-1 dup scale}\box#1\vss}\hss}\@rotfinish}%
\def\@rotf#1{\hbox to\wd#1{\hskip\wd#1\@rotstart{-1 1 scale}%
   \box#1\hss}\@rotfinish}%
\def\rotate{\@ifnextchar[{\@rotate}{\@rotate[l]}}
\def\@rotate[#1]#2{\setbox\@rotbox=\hbox{#2}\@nameuse{@rot#1}\@rotbox}

\catcode`\@=12

\begin{document}

\makeatletter \@addtoreset{equation}{section} \makeatother
\renewcommand{\theequation}{\thesection.\arabic{equation}}
\pagestyle{empty}
\pagestyle{empty}
\vspace{0.5in}
\rightline{FTUAM-02/25}
\rightline{IFT-UAM/CSIC-02-44}
\rightline{\today}
\vspace{2.0cm}
\setcounter{footnote}{0}

\begin{center}
\LARGE{
{\bf Deformed Intersecting D6-Brane GUTS II}}
\\[4mm]
{\large{ Christos ~Kokorelis \footnote{ Christos.Kokorelis@uam.es} }
\\[1mm]}
\normalsize{\em Departamento de F\'\i sica Te\'orica C-XI and 
Instituto de F\'\i sica 
Te\'orica C-XVI}
,\\[-0.3em]
{\em Universidad Aut\'onoma de Madrid, Cantoblanco, 28049, Madrid, Spain}
\end{center}
\vspace{1.0mm}


\begin{center}
{\small  ABSTRACT}
\end{center}
By employing D6-branes
intersecting at angles in $D = 4$ type I strings, 
we construct {\em five stack} string GUT
models (PS-II class), that contain 
at low energy {\em exactly the Standard model}
 with no extra matter and/or
extra gauge group factors.
These classes of models are based on the Pati-Salam (PS)
gauge group
$SU(4)_C \times SU(2)_L \times SU(2)_R$.
They represent deformations around the quark and lepton
basic intersection number structure.

The models possess the same phenomenological
characteristics
of some recently discussed examples (PS-A and PS-I
class) of
four stack
PS GUTS. Namely,
there are no colour triplet couplings to mediate proton decay
 and proton is stable
 as baryon number is a gauged symmetry. Neutrinos get masses
 of the correct sizes. Also the mass relation
 $m_e = m_d$ at the GUT scale is recovered.
 The conditions for the non-anomalous U(1)'s to
 survive
 massless the
 Green-Schwarz mechanism are equivalent, to the
 conditions, coming from the presence
 of N=1 supersymmetry, in sectors involving
 the presence of {\em extra}
 branes and also required to guarantee the
 existence of the Majorana mass
 term for the right handed neutrinos. These conditions are
 independent from the number of {\em extra} U(1) branes.
 We also discuss the relative size of the leading
 worldsheet instanton correction to the  
 trilinear Yukawa couplings in a general GUT
 model.

\newpage

\setcounter{page}{1} \pagestyle{plain}
\renewcommand{\thefootnote}{\arabic{footnote}}
\setcounter{footnote}{0}

\section{Introduction}

 The purpose of this paper is 
to
present three generation five stack string GUT models
that break at low energy exactly to
the Standard Model (SM), 
$SU(3)_C \t SU(2)_L \t U(1)_Y$,
without any extra chiral fermions and/or extra gauge group factors.  
The four-dimensional models are non-supersymmetric
and are based on the Pati-Salam (PS)
$SU(4)_C \times SU(2)_L \times SU(2)_R$ gauge
group.
The basic structure behind the models includes
D6-branes intersecting each other at non-trivial angles,
in an orientifolded compactification of IIA theory on a
factorized six-torus,
where O$_6$ orientifold planes are on top of D6-branes
\cite{tessera, tessera2}.
\newline
The proposed classes of models have some distinctive features :
\begin{itemize}

\item   The models, from now on characterized as belonging
to
the PS-II class,
are being build with a
gauge group $U(4) \times
U(2) \times U(2) \times U(1) \times U(1)$ at the string
scale. At
the scale of
symmetry
breaking of the left-right symmetry \footnote{could be
as high as the
string scale}, $M_{GUT}$, the initial symmetry group
breaks to the the standard model  
$SU(3)_C \times SU(2)_L \times U(1)_Y $
 augmented with two extra anomaly free $U(1)$ symmetries.
 The additional $U(1)$'s may break by the vev of
 some charged singlet
 scalars, e.g. $s^1_B$, $s^2_B$
 at
 a scale set by their vevs, leaving at low energies
 the SM itself.
 The fermions get charged under the broken
 $U(1)$ symmetry, acquiring a flavour symmetry.
 The singlets responsible for breaking the
 $U(1)$ symmetries are obtained by
 demanding certain open sectors
 to respect
 $N=1$ supersymmetry, e.g. $dd^{\star}$, $ee^{\star}$.

\item  A numbers of extra U(1)'s added to cancel the
RR tadpoles results in scalar singlet generation
in combination
with preserving N=1 SUSY on intersections.

\item  Neutrinos gets a mass 
 of the right order, consistent with the LSND oscillation 
experiments, 
from a see-saw mechanism, where the Dirac and Majorana
terms are, of the Frogatt-Nielsen type.

\item  Proton is stable due to the fact that baryon
number is an
unbroken gauged 
global symmetry surviving at low energies and no colour
triplet
couplings that could mediate
proton decay exist. 
Thus a gauged baryon number provides a natural
explanation for
proton stability.
As in the other
D6-models \cite{louis2, kokos, kokos2, kokos1},
on the same background
with just the SM at low
energy \footnote{We note that
there are also intersecting D5-brane 
constructions \cite{D5} \cite{D51} with
exactly the SM at low energy. They will be mentioned
later in this section.},
the baryon number associated
$U(1)$ gauge boson becomes massive through its
couplings to Green-Schwarz
mechanism. That has an an immediate effect that
baryon number is surviving
as a global symmetry to low energies providing for a
natural explanation
for proton stability in general brane-world scenarios.

\item
The model uses small Higgs representations in the adjoint
to break the PS symmetry,
instead of using large
Higgs representations \footnote{ E.g. 126 like in the
standard $SO(10)$ models.}.

\end{itemize}

Some of the major problems of string 
theory is the hierarchy of scale and 
particle masses after 
supersymmetry breaking. 
These phenomelogical issues have by far been explored
in the context of
construction of semirealistic supersymmetric models 
of 
weakly coupled $N=1$ (orbifold) compactifications of the 
heterotic string theories. In these theories 
many problems remain unsolved, we mention briefly one of
them, namely that
the string scale which is of the order of $10^{18}$ GeV
is in clear disagreement with the observed
unification of
gauge
coupling constants in the MSSM of $10^{16}$ GeV.
The latter problem remaind unclear 
even though
the observed discrepancy between the two high 
scales was attributed \footnote{among other options,} to
the presence of the $N=1$
string threshold corrections
to the gauge coupling constants \cite{dikomou}.
Also semirealistic model building has by far
been explored in the context of orientifold
models \cite{dio1}.

On the contrary in type I models, the string scale, which 
is a free parameter, can be lowered in the
TeV range \cite{antoba} thus suggesting 
that non-SUSY models with a string scale in the TeV region
is a viable possibility.
In this spirit, recently some new constructions have
appeared in a type I
string vacuum background which use 
intersecting branes \cite{tessera} and give four
dimensional non-supersymmetric models.\newline
Hence, by using
background fluxes
in a D9 brane type I background \footnote{
In the T-dual 
language these backgrounds are represented by
D6 branes wrapping 3-cycles on a dual torus
and intersecting each other at certain angles. }
it was possible for open sting modes to
be formulated \cite{tessera} that
break supersymmetry on the 
brane and give chiral fermions 
with an even number of 
generations \cite{tessera}.
In these models the fermions get localized
in the
intersections between
branes \cite{bele}. With the
introduction of a quantized background NS-NS B
field \cite{eksi1,eksi2,eksi3}, that makes the tori tilted, is was then
it was then possible to  
give rise to 
semirealistic models with 
three generations \cite{tessera2}.
It should be noted that these backgrounds
are T-dual to models with magnetic
deformations \cite{carlo}.

Furthermore, an important step was taken
in \cite{louis2},
by showing how to construct
the standard model (SM) spectrum together with
right handed
neutrinos in a systematic way. The authors considered,
as a
starting point,
IIA theory compactified on $T^6$ \cite{tessera}
assigned with an
orientifold
product $\Omega \times R$, where $\Omega$ is the
worldsheet parity
operator
and R is the reflection operator with respect to one of the axis of
each tori.
In this case, the four stack D6-branes contain Minkowski space and  
each of the 
three remaining dimensions is wrapped up on a different $T^2$ torus.
In this
construction the proton is stable since the baryon number
is a gauged
$U(1)$ global symmetry.
A special feature of these models is that the neutrinos
can only get Dirac mass. These models have been
generalized to models with
five stacks \cite{kokos} and six stacks 
of D6-branes at the string scale \cite{kokos2}.
The models of \cite{kokos} \cite{kokos2} are build as
deformations of the QCD intersection numbers, namely
they are build around the left and right handed quarks
intersection numbers. Also, they hold exactly the same
phenomenological properties of \cite{louis2}. They also,
have a special feature since by demanding
the presence of $N=1$ supersymmetric sectors, we are
able to break the extra, beyond the SM gauge group,
$U(1)$'s, and thus predicting the unique existence of one
supersymmetric partner of the right neutrino or
two supersymmetric partners of the right neutrinos in
the five and the six stack SM's respectively.

In addition, in \cite{kokos1} we presented the first
examples of GUT models in a string theory context,  and 
in the context of intersecting
branes, that break completely to the SM at
low energies. The models predict uniquely the existence
of light weak fermion doublets with energy between the
range 90 - 246 GeV, that is they can be directly tested at
present of future accelerators. Deformations of these models
will be pursued in this work.

We note that apart from D6 models with exactly the SM at low
energy just mentioned, there are studies using intersecting
D5-branes and only the SM at low energy \cite{D5, D51}. 
In the latter models \cite{D51} there are special classes of theories,
again appearing as deformations of the QCD intersection structure, which have 
not only the SM at low energy but exactly the same low energy effective 
theory including fermion and scalar spectrum.

Also non-SUSY and SUSY constructions in the context of
intersecting branes with the SM plus additional
massless exotic matter were considered in
\cite{luis1, uran}.
For constructions with intersecting branes on compact
Calabi-Yau
spaces see \cite{cala} and for intersecting branes
on non-compact
Calabi-Yau
spaces see \cite{uran1}.
For some other work in the context of intersecting
branes see \cite{alda, cim, cim1, nano, maria, bkl}.

The paper is organized as follows. 
 In section two 
 we describe the general rules for
 building chiral GUT models in orientifolded $T^6$
 compactifications and the possible open string
 sectors. In section 3, we discuss the basic
 fermion and scalar structure of the PS-II class
 of models that will mainly focus in
 this work.
 In section 4,
 we
discuss the parametrization of the solutions to the
RR tadpole cancellation conditions.
In section 5 we discuss the cancellation of $U(1)$
anomalies in
the presence of a 
generalized 
Green-Schwarz (GS) mechanism
and extra $U(1)$ branes.
In section 6, we discuss the conditions for the
absence of tachyons as well describing the
 Higgs sector
of the models including the low energy and high
energy Higgs
sector of the classes of GUTS presented.
In subsection 7.1 we discuss the importance of creating
sectors preserving N=1 SUSY for the realization of the
see-saw mechanism.
In subsection 7.2 we discuss in which way by adding
extra U(1) branes we create scalar
singlets and satisfy RR tadpole conditions.
In subsection 7.3 we discuss the breaking of the
surviving the Green-Schwarz mechanism massless $U(1)$'s
with the use of singlets coming the non-trivial
intersections of the extra branes and leptonic branes
as well
from sectors in the form $jj^{\star}$.
In subsection 8.1 we examine the structure of GUT
Yukawa
couplings in intersecting braneworlds and compute the
leading
worldsheet corrections for the present models.
In subsection 8.2
we examine the problem of
neutrino masses.
In subsection 8.3
we show that all
additional exotic fermions beyond those of SM present in the
models become
massive
and disappear from the low energy spectrum.
Section 9 contains our conclusions. Finally, 
Appendix I, includes the conditions for the
absence of tachyonic modes
in the spectrum of the PS-II class of models presented,
In Appendix II, we discuss the importance of choosing appropriate 
locations for the extra 
branes across the three-cycles, such that the models realize the presence 
of Higgses, needed for
electroweak symmetry breaking.

\section{\em Tadpole structure and spectrum rules}

Next, we describe the construction of the PS classes of models. It is 
based on  
type I string with D9-branes compactified on a six-dimensional
orientifolded torus $T^6$, 
where internal background 
gauge fluxes on the branes are turned on \cite{bachas, tessera, tessera2}. 
By performing a T-duality transformation on the $x^4$, $x^5$, $x^6$, 
directions the D9-branes with fluxes are translated into D6-branes 
intersecting at 
angles. The branes are not paralled to the
orientifold planes.
We assume that the D$6_a$-branes are wrapping 1-cycles 
$(n^i_a, m^i_a)$ along each of the $T^2$
torus of the factorized $T^6$ torus, namely 
$T^6 = T^2 \times T^2 \times T^2$.

In order to build a PS model with minimal Higgs
structure we consider
four stacks of D6-branes giving rise to their world-volume to an initial
gauge group $U(4) \times U(2) \times U(2) \times U(1) \times U(1) $
at the string
scale.
In addition, we consider the addition of NS B-flux, 
such that the tori are not orthogonal,
avoiding in this way an even number of families, 
and leading to effective tilted wrapping numbers, 
\beq
(n^i, m ={\tilde m}^i + {n^i}/2);\; n,\;{\tilde m}\;\in\;Z,
\label{na2}
\eeq
that allows semi-integer values for the m-numbers.  
\newline
In the presence of $\Omega {\cal R}$ symmetry, 
where $\Omega$ is the worldvolume 
parity and $\cal R$ is the reflection on the T-dualized coordinates,
\beq
T (\Omega {\cal R})T^{-1}= \Omega {\cal R},
\label{dual}
\eeq
and thus each D$6_a$-brane
1-cycle, must have its $\Omega {\cal R}$ partner $(n^i_a, -m^i_a)$. 

Chiral fermions are obtained by stretched open strings between
intersecting D6-branes \cite{bele}. 
Also the chiral spectrum of the models may be obtained 
after solving simultaneously 
the intersection 
constraints coming from the existence of the different sectors together
with the RR
tadpole cancellation conditions.

There are a number of different sectors, 
which should be taken into account when computing the chiral spectrum.
We will denote the action of 
$\Omega R$ on a sector $\a, \b$, by ${\a}^{\star}, {\b}^{\star}$,
respectively.
The possible sectors are:

\begin{itemize}
 
\item The $\a \b + \b \a$ sector: involves open strings stretching
between the
D$6_{\a}$ and D$6_{\b}$ branes. Under the $\Omega R$ symmetry 
this sector is mapped to its image, ${\a}^{\star} {\b}^{\star}
+ {\b}^{\star} {\a}^{\star}$ sector.
The number, $I_{\a\b}$, of chiral fermions in this sector, 
transforms in the bifundamental representation
$(N_{\a}, {\bar N}_{\a})$ of $U(N_{\a}) \times U(N_{\b})$, and reads
\beq
I_{\a\b} = ( n_{\a}^1 m_{\b}^1 - m_{\a}^1 n_{\b}^1)( n_{\a}^2 m_{\b}^2 -
m_{\a}^2 n_{\b}^2 )
(n_{\a}^3 m_{\b}^3 - m_{\a}^3 n_{\b}^3),
\label{ena3}
\eeq
 where $I_{\a\b}$
is the intersection number of the wrapped cycles. Note that the sign of
 $I_{\a\b}$ denotes the chirality of the fermion and with $I_{\a\b} > 0$
we denote left handed fermions.
Negative multiplicity denotes opposite chirality.

\item The $\a\a$ sector : it involves open strings stretching on a single 
stack of 
D$6_{\a}$ branes.  Under the $\Omega R$ symmetry 
this sector is mapped to its image ${\a}^{\star}{\a}^{\star}$.
 This sector contain ${\cal N}=4$ super Yang-Mills and if it exists
SO(N), SP(N) groups appear. 
This sector is of no importance to us as we will be dealing with
 unitary groups.

\item The ${\a} {\b}^{\star} + {\b}^{\star} {\a}$ sector :
It involves chiral fermions transforming into the $(N_{\a}, N_{\b})$
representation with multiplicity given by
\beq
I_{{\a}{\b}^{\star}} = -( n_{\a}^1 m_{\b}^1 + m_{\a}^1 n_{\b}^1)
( n_{\a}^2 m_{\b}^2 + m_{\a}^2 n_{\b}^2 )
(n_{\a}^3 m_{\b}^3 + m_{\a}^3 n_{\b}^3).
\label{ena31}
\eeq
Under the $\Omega R$ symmetry transforms to itself.

\item the ${\a} {\a}^{\star}$ sector : under the $\Omega R$ symmetry is 
transformed to itself. From this sector the invariant intersections
will give 8$m_{\a}^1 m_{\a}^2 m_{\a}^3$ fermions in the
antisymmetric representation
and the non-invariant intersections that come in pairs provide us with
4$ m_{\a}^1 m_{\a}^2 m_{\a}^3 (n_{\a}^1 n_{\a}^2 n_{\a}^3 -1)$ additional 
fermions in the symmetric and 
antisymmetric representation of the $U(N_{\a})$ gauge group.

\end{itemize}

Also any vacuum derived from the previous intersection number constraints of the 
chiral spectrum 
is subject to constraints coming from RR tadpole cancellation 
conditions \cite{tessera}. That requires cancellation of 
D6-branes charges \footnote{Taken together with their
orientifold images $(n_a^i, - m_a^i)$  wrapping
on three cycles of homology
class $[\Pi_{\alpha^{\star}}]$.}, wrapping on three cycles with
homology $[\Pi_a]$ and O6-plane 7-form
charges wrapping on 3-cycles with homology $[\Pi_{O_6}]$. In formal terms,
the RR tadpole cancellation  conditions
in terms of cancellations of RR charges in homology, read :
\beq
\sum_a N_a [\Pi_a]+\sum_{\alpha^{\star}} 
N_{\alpha^{\star}}[\Pi_{\alpha^{\star}}] -32
[\Pi_{O_6}]=0.
\label{homology}
\eeq  
Explicitly, the RR tadpole conditions read :
\beqa
\sum_a N_a n_a^1 n_a^2 n_a^3 =\ 16,\nonumber\\
\sum_a N_a m_a^1 m_a^2 n_a^3 =\ 0,\nonumber\\
\sum_a N_a m_a^1 n_a^2 m_a^3 =\ 0,\nonumber\\
\sum_a N_a n_a^1 m_a^2 m_a^3 =\ 0.
\label{na1}
\eeqa
That ensures absence of non-abelian gauge anomalies.
A comment is in order. It is important to notice
that the RR tadpole cancellation condition can be understood as
a constraint that demands that for each gauge group the number of
fundamentals to be equal to the number of bifundamendals.
As a general rule to D-brane model building, by considering $a$ 
stacks of D-brane configurations with 
$N_a, a=1,.., N$, paralled branes, the gauge group that appears is in 
the form $U(N_1) \times U(N_2) \times \cdots \times U(N_a)$. Thus,
effectively each $U(N_i)$ factor will give rise to an $SU(N_i)$,
charged under the associated $U(1_i)$ gauge group factor, that appears in 
the decomposition $SU(N_a) \times U(1_a)$.
A brane configuration with the unique minimal PS particle content
such that intersection numbers, tadpole conditions and various phenomenological
requirements including the presence of exotic representations 
are accommodated, 
 can be 
obtained by considering five stacks of branes yielding an initial
$U(4)_a \times U(2)_b \times U(2)_c \times U(1)_d \times U(1)_e $. In this 
case the equivalen
gauge group is an $SU(4)_a \times SU(2)_b \times SU(2)_b 
\times U(1)_a \times U(1)_b 
\times U(1)_c \times U(1)_d \times U(1)_e$. Thus, in the first instance, 
we can identify, without loss of 
generality, $SU(4)_a$ as the $SU(4)_c$ colour group 
that its breaking could induce 
the $SU(3)$ colour group of strong interactions, the $SU(2)_b$ with 
$SU(2)_L$ of weak interactions and $SU(2)_c$ with $SU(2)_R$  of left-right 
symmetric PS models.

\section{\em The basic fermion  structure}

The basic PS-II class of models that we will center our attention
in this work, will be a three family non-supersymmetric
GUT model with the left-right
symmetric Pati-Salam model structure \cite{pati}
$SU(4)_C \times SU(2)_L \times SU(2)_R$.
The open string background on which the models will
be build will be intersecting D6-branes wrapping
on 3-cycles of
decomposable toroidal ($T^6$) orientifolds of type IIA
in four dimensions \cite{tessera, tessera2}.
 
The three generations of quark and lepton 
fields are accommodated into the following
representations :
\beqa
F_L &=& (4, {\bar 2}, 1) =\ q(3, {\bar 2}, \frac{1}{6})
+\ l(1, {\bar 2}, -\frac{1}{2})
\equiv\ (\ u,\ d,\ l), \nonumber\\
{\bar F}_R &=& ({\bar 4}, 1,  2) =\ {u}^c({\bar 3}, 1, - \frac{2}{3}) +
{d}^c({\bar 3}, 1, \frac{1}{3}) + {e}^c(1, 1, 1) + {N}^c(1, 1, 0) 
\equiv ( {u}^c,  {d}^c,   {l}^c),\nonumber\\ 
\label{na3}
\eeqa
where the quantum numbers on the right hand side of 
(\ref{na3})
are with respect to the decomposition of the $SU(4)_C \times SU(2)_L \times 
SU(2)_R$ under the $SU(3)_C \times SU(2)_L \times U(1)_Y$ gauge group and
${l}=(\nu,e) $ is the standard left handed lepton doublet,
 ${l}^c=({N}^c, e^c)$ are the right handed leptons.
Also the assignment of the accommodation of the quarks and leptons into
the representations $F_L + {\bar F}_R$ is the one appearing in the
spinorial decomposition of the $16$ representation
of $SO(10)$ under the PS gauge group.
\newline
A set of useful fermions appear also in the model
\beq
\chi_L =\ ( 1,  {\bar 2}, 1),\ \chi_R =\ (1, 1, {\bar 2}).
\label{useful}
\eeq
These fermions are a general prediction
of left-right symmetric theories as the existence
of these representations
follows from RR tadpole cancellation conditions. 
\newline
The symmetry breaking of the left-right PS symmetry
at the $M_{GUT}$ scale \footnote{In principle
this scale could be as high as the
string scale.}
proceeds through the representations of
the set of Higgs fields,
\beq
H_1 =\ ({\bar 4}, 1, {\bar 2}), \
H_2 =\ ( 4, 1, 2),
\label{useful1}
\eeq
 where,
\beqa
H_1 = ({\bar 4}, 1, {\bar 2}) = {u}_H({\bar 3}, 1, \frac{2}{3}) +
{d}_H({\bar 3}, 1, -\frac{1}{3}) + {e}_H(1, 1, -1)+
{\nu}_H(1, 1, 0).
\label{higgs1}
\eeqa
The electroweak symmetry breaking is delivered 
through bi-doublet Higgs fields $h_i$ $i=3, 4$,
field in the representations 
\beq
h_3 =\ (1, {\bar 2}, 2),\ h_4 =\ (1, 2, {\bar 2}) \ .
\label{additi1}
\eeq
Because of the imposition of N=1 SUSY on some open string
sectors, there are also present
 the massless scalar superpartners of the quarks, leptons and
antiparticles 
\beq
{\bar F}_R^H = ({\bar 4}, 1,  2) = {u}^c_H({\bar 3}, 1, -\frac{4}{6})+
{d}^c_H({\bar 3}, 1, \frac{1}{3})+ {e}^c_H(1, 1, 1) + 
{N}^c_H(1, 1, 0) 
\equiv ({u}^c_H, {d}^c_H,  {l}^c_H).\nonumber\\
\label{na368}
\eeq
The latter fields \footnote{
are replicas
of the fermion fields appearing in the intersection 
$ac$ and they receive a vev} characterize all vacua
coming from these type IIA orientifolded tori 
constructions is the replication of massless fermion spectrum
by an equal number of massive particles in the
same representations and with the same quantum numbers.
\newline
This is the basic fermionic structure appearing in 
the PS models that we
have considered in \cite{kokos1} and will be appearing
later in this work.
Also, a number of charged exotic
fermion fields, which receive a string scale mass, appear
\beq
6(6, 1, 1),\;\; \ 6({\bar 10}, 1, 1).
\label{beg1}
\eeq

The complete accommodation of the fermion structure of
the PS-II classes of models under study in this work can be seen
in table  (\ref{spectrum8}).

\begin{table}[htb] \footnotesize
\renewcommand{\arraystretch}{1.5}
\begin{center}
\begin{tabular}{|c|c||c|c||c||c|c|c|}
\hline
Fields &Intersection  & $\bullet$ $SU(4)_C \times SU(2)_L \times SU(2)_R$
 $\bullet$&
$Q_a$ & $Q_b$ & $Q_c$ & $Q_d$ & $Q_e$\\
\hline
 $F_L$& $I_{ab^{\ast}}=3$ &
$3 \times (4,  2, 1)$ & $1$ & $1$ & $0$ &$0$ &$0$\\
 ${\bar F}_R$  &$I_{a c}=-3 $ & $3 \times ({\ov 4}, 1, 2)$ &
$-1$ & $0$ & $1$ & $0$ & $0$\\
 $\chi_L^1$& $I_{bd^{\star}} = -8$ &  $8 \times (1, {\ov 2}, 1)$ &
$0$ & $-1$ & $0$ & $-1$ & $0$\\    
 $\chi_R^1$& $I_{cd} = -8$ &  $8 \times (1, 1, {\ov 2})$ &
$0$ & $0$ & $-1$ &$1$ &$0$\\
 $\chi_L^2$& $I_{be} = -4$ &  $4 \times (1, {\ov 2}, 1)$ &
$0$ & $-1$ & $0$ &$0$ & $1$ \\    
 $\chi_R^2$& $I_{ce^{\ast}} = -4$ &  $4 \times (1, 1, {\ov 2})$ &
$0$ & $0$ & $-1$ & $0$ &$-1$ \\\hline
 $\omega_L$& $I_{aa^{\ast}}$ &  $6 \b^2
 \times (6, 1, 1)$ & $2$ & $0$ & $0$ &$0$ &$0$\\
 $y_R$& $I_{aa^{\ast}}$ & $6  \b^2  \times ({\bar 10}, 1, 1)$ &
$-2$ & $0$ & $0$ &$0$ &$0$ \\
\hline
 $s_R^1$ & $I_{dd^{\ast}}$ &  $16 \b^2
 \times (1, 1, 1)$ & $0$ & $0$ & $0$ &$2$ &$0$\\
 $s_R^2$ & $I_{ee^{\ast}}$ & $8  \b^2  \times (1, 1, 1)$ &
$0$ & $0$ & $0$ &$0$ &$-2$ \\
\hline
\end{tabular}
\end{center}
\caption{\small Fermionic spectrum of the $SU(4)_C \times
SU(2)_L \times SU(2)_R$, PS-II class of models together with $U(1)$
charges. We note that at energies of order $M_z$ only
the Standard model survives.
\label{spectrum8}}
\end{table}

\section{\em Tadpole cancellation for PS-II classes of GUTS}

To understand the solution of the RR tadpole cancellation condition, that it
will be given in parametric form, we 
should make the following comments :
\newline
a) The need to 
realize certain couplings will force us to demand that some 
intersections will
preserve some supersymmetry. Thus some massive
fields will be
``pulled out" from the massive spectrum and become massless.
For example, in order to realize a Majorana mass term for the right
handed neutrinos we will demand that the sector
$ac$ preserves $N=1$ SUSY. That will have as an
immediate effect
to "pull out" from the massive mode spectrum the ${\bar F}_R^H$ particles. 
\newline
b)
The intersection numbers, in table (\ref{spectrum8}),  
of the fermions $F_L + {\bar F}_R$ are chosen 
such that $I_{ac} = - 3$, $I_{ab^{\star}} = 3$. Here, $-3$ denotes opposite 
chirality to that of a left handed fermion. 
The choice of additional fermion representations
$(1, {\bar 2} ,1)$,
$(1, 1, {\bar 2})$ is imposed to us by the RR tadpole cancellation conditions
that are equivalent to
$SU(N_a)$ gauge
anomaly cancellation, in this case of $SU(2)_L$,
$SU(2)_R$ gauge anomalies,
\beq
\sum_i I_{ia} N_a = 0,\;\;a = L, R.
\label{ena4}
\eeq
c)
The PS-II class of models don't accommodate
representations
of scalar sextets $(6, 1, 1)$ fields,
that appear in attempts to construct realistic
4D $N=1$ PS
heterotic models
from
the fermionic formulation \cite{antoI},
even 
through heterotic fermionic models where those 
representations are lacking 
exist \cite{giapo}.
Those representations were imposed earlier in
attempts to produce a
realistic PS model
as a recipe for saving the models from proton decay.
Fast proton decay was avoided by making 
the mediating 
$d_H$ triplets of (\ref{higgs1}) superheavy 
and of order of the $SU(2)_R$ breaking scale
via their couplings to the sextets.
In the models we examine in this work,
baryon number is a gauged global symmetry,
so that proton is 
stable. Thus there is no need to introduce sextets to save the models
from fact proton decay as proton is stable.
\newline
Also in the present PS-II GUTS, there is no problem of having
$d_H$ becoming light
enough and causing catastrofic proton decay, as the only way this could
happen, is through the existence of the $d_H$
coupling to sextets
to quarks and leptons. However, this coupling is forbidden by
the symmetries 
of the models.
\newline
 The theory breaks
just to the standard model $SU(3) \times SU(2) \times U(1)_Y$
at low energies. The tadpole solutions
of PS-II
models are presented in table (\ref{spectruma101}).
\newline
d) The mixed anomalies $A_{ij}$ of the seven \footnote{We examine for 
convenience the case of two added extra U(1)'s.}
surplus $U(1)$'s
with the non-abelian gauge groups $SU(N_a)$ of the theory
cancel through a generalized GS mechanism \cite{sa,iru, louis2, kokos,
kokos2},
involving
 close string modes couplings to worldsheet gauge fields.
 Two combinations of the $U(1)$'s are anomalous and become
 massive through their
 couplings to RR fields, their
 orthogonal 
 non-anomalous combinations survives, combining to a
 single $U(1)$
 that remains massless. Crucial for achieving the RR tadpole
 cancellation is the presence of $N_h$ extra branes. Contrary,
 of what
 is happening in D6-brane models \footnote{Also happening in
 intersecting D5-brane models, with exactly the SM at low energy
  and a Standard-like structure at the string scale \cite{D5, D51}.},
  with exactly the SM at low energy,
  and a Standard-like structure at the string
  scale \cite{louis2, kokos, kokos2} where
  the extra branes have no intersection with the branes,
  in the intersecting GUT models there is a non-vanishing intersection of the 
extra branes with the rest of the branes.
  As a consequence, this becomes a singlet generation
  mechanism after imposing $N=1$ SUSY between $U(1)$
   leptonic (the $d$, $e$ branes) and the 
  $U(1)$ extra branes.
  Also, contrary to the SM's of
  \cite{louis2, kokos, kokos2, D5, D51} the extra branes
  do not form a $U(N_h)$ gauge group but rather a
   $U(1)^{N_1} \times U(1)^{N_2} \cdots U(1)^{N_h}$ one.
\newline
e)
The constraint 
\beqa
{\Pi}_{i=1}^3 m^i&=&0.
\label{req1}
\eeqa                  
is not imposed 
and thus
leads to the appearance of the non-trivial chiral
fermion content from the
$aa^{\ast}$, $dd^{\ast}$, $ee^{\ast}$ sector with corresponding 
fermions $\omega_L$, $z_R$, $s_R^1$, $s_R^2$.\newline
f)
After breaking the PS left-right symmetry at $M_{GUT}$, the
surviving gauge symmetry
is that of the SM augmented by four anomaly free $U(1)$ symmetries, 
including the added extra U(1) branes,
surviving the Green-Schwarz mechanism.  
To break the latter $U(1)$ symmetries we will impose
that the $dd^{\star}$, $ee^{\star}$$dh$, $dh^{\star}$ sectors 
\footnote{WE denoted by $h$ the presence of extra branes.} 
respects $N=1$ SUSY.
Thus singlets scalars will appear, that are
superpartners of the corresponding fermions.
\newline
f) Demanding $I_{ab^{\star}}=3$, $I_{ac}=-3$,
it implies that the third tori should be tilted. By looking at the 
intersection numbers of table one,  we conclude that the
b-brane should be paralled to the c-brane and the a-brane should be 
paralled to the d, e branes as there is an absence of intersection 
numbers for 
those branes.
The  
cancellation of the RR crosscap tadpole constraints
is solved from multiparametric sets of solutions which are given 
in table (\ref{spectruma101}).

\begin{table}[htb]\footnotesize
\renewcommand{\arraystretch}{2}
\begin{center}
\begin{tabular}{||c||c|c|c||}
\hline
\hline
$N_i$ & $(n_i^1, m_i^1)$ & $(n_i^2, m_i^2)$ & $(n_i^3, m_i^3)$\\
\hline\hline
 $N_a=4$ & $(0, \epsilon)$  &
$(n_a^2, 3 \epsilon {\tilde \epsilon}\b_2)$ & $(1, {\tilde \epsilon}/2)$  \\
\hline
$N_b=2$  & $(-1, \epsilon m_b^1 )$ & $(1/\beta_2, 0)$ &
$(1, {\tilde \epsilon}/2)$ \\
\hline
$N_c=2$ & $(1, \epsilon m_c^1 )$ &   $(1/\beta_2, 0)$  & 
$(1, -{\tilde \epsilon}/2)$ \\    
\hline
$N_d=1$ & $(0, \epsilon)$ &  $(n_d^2, -4 \epsilon
{\tilde \epsilon}\b_2)$
  & $(2, {\tilde \epsilon})$  \\\hline
$N_e=1$ & $(0, \epsilon)$ &  $(n_e^2, -2 \epsilon
{\tilde \epsilon}\b_2)$
  & $(2, -{\tilde \epsilon})$  \\\hline
$1$& $(1/\beta_1, 0)$ &  $(1/\beta_2, 0)$
  & $(2, 0)$  \\\hline
$\vdots$& $\vdots$ &  $\vdots$
  & $\vdots$  \\\hline
$N_h$ & $(1/\beta_1, 0)$ &  $(1/\beta_2, 0)$
  & $(2, 0)$  \\\hline   
\end{tabular}
\end{center}
\caption{\small
Tadpole solutions for PS-II type models where the five stack of 
D6-branes wrapping numbers giving rise to the 
fermionic spectrum and the SM,
$SU(3)_C \times SU(2)_L \times U(1)_Y$, gauge group at low energies.
The wrappings 
depend on two integer parameters, 
$n_a^2$, $n_d^2$, $n_e^2$, the NS-background $\beta_i$ and the 
phase parameters $\epsilon = {\tilde \epsilon }= \pm 1$. 
Also there is an additional dependence on the two wrapping
numbers, integer of half integer,
$m_b^1$, $m_c^1$. Note also that the presence of the
$N_h$ extra U(1) branes.
\label{spectruma101}}
\end{table}

A comment is in order.
The location of extra branes needed to satisfy
the RR tadpoles is particularly important.
Choosing e.g. $\beta_1 = \beta_2 =1/2$ and
 their location to be at
\beq
(1/\beta^1, 0) (1/\beta^1, 0)(1, m/2), \ m \in 2 Z + 1
\label{loc1}
\eeq
it results in classes of models with no electroweak
bidoublets $h_1$, $h_2$, $h_3$, $h_4$.
The relevant analysis can be seen in Appendix II.
Thus we choose to add $N_h$ extra branes
located at
\beq
(1/\beta_1, 0)(1/\beta_2, 0)(2, 0)
\label{2ndform}
\eeq
In principle the extra branes could have
a different from but great care should be taken,
as the non-zero intersections of the extra branes
with the rest of the branes could create massless exotic
fermions that cannot be made massive and disappear from the
low energy spectrum.

The first tadpole condition in (\ref{na1}) 
depends on the number of extra branes that it is added.
Thus it becomes 
\beq
N_h \frac{2}{\beta_1 \beta_2} = 16.
\eeq
The third tadpole condition becomes
\beq
(2 n_a^2 + n_d^2 - n_e^2)
+ \frac{1}{\beta_2}(m_b^1 -m_c^1) =\ 0.
\label{ena11}
\eeq

To see clearly the cancellation of tadpoles, we have
to choose a
consistent numerical set of wrapping
numbers, e.g.
\beq
\epsilon =\ {\tilde \epsilon} =\ 1,\ n_a^2=1,
\;m_b^1=-3/2,\ m_c^1= 1/2, \
n_d^2=1,\ n_e^2=1,
\;\b_1=1/2, \  \b_2=1.
\label{numero1}
\eeq
The latter can be satisfied with the addition of four
extra U(1)
D6-branes.
\begin{table}[htb]\footnotesize
\renewcommand{\arraystretch}{2}
\begin{center}
\begin{tabular}{||c||c|c|c||}
\hline
\hline
$N_i$ & $(n_i^1, m_i^1)$ & $(n_i^2, m_i^2)$ & $(n_i^3, m_i^3)$\\
\hline\hline
 $N_a=4$ & $(0, 1)$  &
$(1, 3)$ & $(1, 1/2)$  \\
\hline
$N_b=2$  & $(-1, -3/2 )$ & $(1, 0)$ &
$(1, 1/2)$ \\
\hline
$N_c=2$ & $(1, 1/2 )$ &   $(1, 0)$  & 
$(1, -1/2)$ \\    
\hline
$N_d=1$ & $(0, 1)$ &  $(1, -4)$
  & $(2, 1)$  \\   
\hline
$N_e=1$ & $(0, 1)$ &  $(1, -2)$
  & $(2, -1)$  \\   
\hline
\end{tabular}
\end{center}
\caption{\small Wrapping number set consistent
with the tadpole constraint (\ref{ena11}). We have not include 
the extra U(1) branes.
\label{new}}
\end{table}

We note that in the model described by
the wrapping numbers of table (\ref{new})
we cannot get the SM at low energy as all the fermions
are charged under the U(1)'s (The U(1)'s can be seen in
 (\ref{PSIIab1}) and at this stage they do not 
remain massless as they have non-zero couplings to RR fields.).

f) the hypercharge operator 
 is defined as usual in this classes of GUT
 models( see also \cite{kokos1}) as
 a linear combination
of the three diagonal generators of the $SU(4)$, $SU(2)_L$, $SU(2)_R$ groups:
\beq
Y = \frac{1}{2}T_{3R}+ \frac{1}{2}T_{B-L},\;T_{3R}=diag(1,-1),\;
T_{B-L}=diag(\frac{1}{3},\frac{1}{3},\frac{1}{3}, -1). 
\label{hyper12}
\eeq 
Also,
\beqa
Q & = & Y   +\ \frac{1}{2}T_{3L}.\\
\label{hye1}
\eeqa

\section{\em Cancellation of U(1) Anomalies}

The mixed anomalies $A_{ij}$ of the four $U(1)$'s
with the non-Abelian gauge groups are given by
\beq
A_{ij}= \frac{1}{2}(I_{ij} - I_{i{j^{\star}}})N_i.
\label{ena9}
\eeq
Moreover, analyzing the mixed anomalies 
of the extra $U(1)$'s with the non-abelian gauge groups $SU(4)_c$, 
$SU(2)_R$, $SU(2)_L$ we can see that there are three anomaly free combinations
$Q_b - Q_c$, $Q_a + Q_d - Q_e$ and $Q_a + 4 Q_d + 5 Q_e$.
Note that gravitational anomalies cancel since D6-branes never 
intersect O6-planes.
In the orientifolded type I torus models gauge anomaly 
cancellation \cite{iru} proceeds through a 
generalized GS
mechanism \cite{louis2} that makes use of the 10-dimensional RR gauge fields
$C_2$ and $C_6$ and gives at four dimensions
the couplings to gauge fields 
 \beqa
N_a m_a^1 m_a^2 m_a^3 \int_{M_4} B_2^o \wedge F_a &;& n_b^1 n_b^2 n_b^3
 \int_{M_4}
C^o \wedge F_b\wedge F_b,\\
N_a  n^J n^K m^I \int_{M_4}B_2^I\wedge F_a&;&n_b^I m_b^J m_b^K \int_{M_4}
C^I \wedge F_b\wedge F_b\;,
\label{ena66}
\eeqa
where
$C_2\equiv B_2^o$ and $B_2^I \equiv \int_{(T^2)^J \times (T^2)^K} C_6 $
with $I=1, 2, 3$ and $I \neq J \neq  K $. Notice the four dimensional duals
of $B_2^o,\ B_2^I$ :
\beqa
C^o \equiv \int_{(T^2)^1 \times (T^2)^2 \times (T^2)^3} C_6&;C^I \equiv
\int_{(T^2)^I} C_2, 
\label{ena7}
\eeqa
where $dC^o =-{\star} dB_2^o,\; dC^I=-{\star} dB_2^I$.

The triangle anomalies (\ref{ena9}) cancel from the existence of the
string amplitude involved in the GS mechanism \cite{sa} in four 
dimensions \cite{iru}. 
The latter amplitude, where the $U(1)_a$ gauge field couples to one
of the propagating
$B_2$ fields, coupled to dual scalars, that couple in turn to
two $SU(N)$ gauge bosons, is 
proportional \cite{louis2} to
\beq
-N_a  m^1_a m^2_a m^3_a n^1_b n^2_b n^3_b -
N_a \sum_I n^I_a n^J_a n^K_b m^I_a m^J_b m^K_b\; ,
I \neq J, K 
\label{ena8}
\eeq

We make the minimal choice
\beq
\beta_1 = \beta_2 =  1/2
\eeq
that requires two extra D6 branes. \newline
In this case
the structure of U(1) couplings reads :
\beqa
B_2^3 \wedge 
[2 {\tilde \epsilon}]
[-(F^b + F^c)],&\nonumber\\
B_2^1 \wedge [\epsilon] 
[ 4 n_a^2 \ F^a +
4 m_b^1 \ F^b + 4 m_c^1 \ F^c + 2 n_d^2  F^d +
2 n_e^2 F^e],&\nonumber\\
B_2^o  \wedge \left(   3  F^a - 2 F^d + F^e  \right) .&
\label{PSIIb}
\eeqa
As can be seen from (\ref{PSIIb}) two anomalous
combinations of $U(1)$'s, e.g.
 $3 F^a - 2 F^d + F^e$, $-(F^b + F^c)$
 become massive through their couplings to RR
 fields $B_2^o$, $B_2^3$. Also there is an anomaly free
 model dependent U(1) which is getting massive from
 its coupling to the RR field $B_2^1$. 
In addition, there are four non-anomalous $U(1)$'s
which also are
getting broken by vevs of singlet scalars generated
by imposing N=1 SUSY on certain sectors.
They are :
\beqa
U(1)^{(4)} =\ (Q^b - Q^c) + (Q^a + Q^d - Q^e),
&\nonumber\\
U(1)^{(5)} =\ F^{{\hat h}_1}, \ \ 
U(1)^{(6)} =\ F^{{\hat h}_2}, &\nonumber\\
U(1)^{(7)} =\ Q^a + 4Q^d +5Q^e \ .
\label{PSIIab1}
\eeqa
The choice of U(1)'s (\ref{PSIIab1}) have no couplings
to RR fields, and thus survive massless the presence of 
the generalized Green-Shwarz mechanism, 
 if  
\beq
2 n_a^2 +\ 4 n_d^2 +\ 5 n_e^2 =\ 0
\label{constr1}
\eeq

At this point we should list 
the couplings of the dual scalars $C^I$ of $B_2^I$
required to cancel
the mixed anomalies of the $U(1)$'s with the 
non-abelian gauge groups $SU(N_a)$.
They are given by
\beqa
C^o \wedge 2  [-(F^b \wedge F^b)
+ (F^c \wedge F^c)], &\nonumber\\
C^2 \wedge [{\epsilon}{\tilde \epsilon}][
  2 n_a^2 (F^a \wedge
F^a)  +  2 m_b^1 (F^b \wedge 
F^b) -  2 m_c^1 (F^c \wedge F^c) 
+ n_d^2 (F^d \wedge F^d) &\nonumber\\
- n_e^2 (F^e
\wedge F^e)],& \nonumber\\
C^3 \wedge  
[\frac{{\tilde \epsilon}}{2}] [ 3(F^a \wedge
F^a)- 8 (F^d \wedge F^d) -4 (F^e \wedge F^e) ],& \nonumber\\
\label{nonanomal}
\eeqa
As it will be shown later, the conditions for demanding that some 
sectors respect N=1 SUSY, that in turn guarantee the existence
of a Majorana coupling for right handed neutrinos as well creating singlets 
necessary to break the U(1)'s (\ref{PSIIab1}), solve the condition
(\ref{constr1}).
We note that if we had chosen $\beta_1 = 1$,
$\beta_2 =1/2$ that is $N_h = 4$ extra branes, we simply
would
have two more U(1) generators surviving massless that
is $U(1)^{(8)} = F^{h3}$, $U(1)^{(9)} = F^{h4}$.
In a similar way we can treat the case
$\beta_1 = \beta_2 =
1$.

\section{Higgs sector, global symmetries, proton stability, $N=1$ SUSY on
intersections and
neutrino masses}

\subsection{\em Stability of the configurations and Higgs sector}

We have so far seen the appearance in the R-sector 
of $I_{ab}$ massless fermions
in the D-brane intersections 
transforming under bifundamental representations $N_a, {\bar N}_b$.
 In intersecting 
brane words, besides the actual
presence of massless fermions at each intersection, 
we have evident the presence of an equal number of
 massive bosons, in the NS-sector, in the same representations 
as the massless fermions \cite{luis1}.
Their mass is of order of the string scale and it should be taken 
into account when examining phenomenological applications related to the
renormalization group equations.
However, it is possible that some of those 
massive bosons may become 
tachyonic \footnote{For consequences
when these set of fields may become massless see \cite{cim}.}, 
especially when their mass, that depends on the 
angles between the branes,
is such that is decreases the world volume of the 
3-cycles involved in the recombination process of joining the two
branes into a single one \cite{senn}.
Denoting the twist vector by $(\vartheta_1,\vartheta_2,
\vartheta_3,0)$, in the NS open string sector the 
lowest lying states are given by \footnote{
we assume $0\leq\vartheta_i\leq 1$ .}
{\small \beqa
\begin{array}{cc}
{\rm \bf State} \quad & \quad {\bf Mass} \\
(-1+\vartheta_1,\vartheta_2,\vartheta_3,0) & \alpha' M^2 =
\frac 12(-\vartheta_1+\vartheta_2+\vartheta_3) \\
(\vartheta_1,-1+\vartheta_2,\vartheta_3,0) & \alpha' M^2 =
\frac 12(\vartheta_1-\vartheta_2+\vartheta_3) \\
(\vartheta_1,\vartheta_2,-1+\vartheta_3,0) & \alpha' M^2 =
\frac 12(\vartheta_1+\vartheta_2-\vartheta_3) \\
(-1+\vartheta_1,-1+\vartheta_2,-1+\vartheta_3,0) & \alpha' M^2
= 1-\frac 12(\vartheta_1+\vartheta_2+\vartheta_3)
\label{tachdsix}
\end{array}
\eeqa}
Exactly at the point, where one of these masses may
become massless we have
preservation of ${\cal N}=1$ SUSY. 
WE note that the angles at the four different intersections can be expressed
in terms of the parameters of the tadpole solutions.

$\bullet$ {\em Angle structure and Higgs fields for PS-II classes of models}

The angles at the different intersections can be expressed in terms of the 
tadpole solution parameters. 
We define the angles:
\beqa
\theta_1 \   = \ \frac{1}{\pi} cot^{-1}\frac{ R_1^{(1)}}{ \epsilon m_b^1 R_2^{(1)}} \ ;\
\theta_2 \  =   \  \frac{1}{\pi} cot^{-1}
\frac{\epsilon {\tilde \epsilon}n_a^2 R_1^{(2)}}{3 \epsilon \beta_2 R_2^{(2)}} \ ;\
\theta_3 \  = \  \frac{1}{\pi} cot^{-1}\frac{2R_1^{(3)}}{R_2^{(3)}},
 \nonumber \\
{\tilde {\theta_1}} \   = 
\ \frac{1}{\pi} cot^{-1}\frac{ R_1^{(1)}}{\epsilon m_c^1 R_2^{(1)}}\
,\ {\tilde {\theta_2}} \   =
\ \frac{1}{\pi} cot^{-1}\frac{\epsilon {\tilde \epsilon}n_d^2 R_1^{(1)}}{ 4
 \b_2 R_2^{(1)}}, \ {\bar {\theta_2}} \   = \
 \frac{1}{\pi} cot^{-1}\frac{\epsilon {\tilde \epsilon}n_e^2 R_1^{(1)}}{ 2
 \beta_2 R_2^{(1)}}
\label{angPSII}
\eeqa
where we consider $\epsilon {\tilde \epsilon} >0$,
$\epsilon m_b^1 >0$, $\epsilon m_c^1 >0$ and 
$R_{i}^{(j)}$, $i={1,2}$ are the compactification radii
for the three $j=1,2,3$ tori, namely
projections 
of the radii 
 onto the cartesian axis $X^{(i)}$ directions when the NS flux B field,
$b^k$, $k=1,2$ is turned on. 

At each of the six non-trivial intersections 
we have the 
presense of four states $t_i , i=1,\cdots, 4$, that could
become massless, associated
to the states (\ref{tachdsix}).
 Hence we have a total of
twenty four different scalars in the model.
The setup is seen clearly if we look at figure one.
These scalars are generally massive but for some values of
their angles could become tachyonic (or massless).

Also, if we demand that the scalars associated with (\ref{tachdsix}) and 
PS-II models 
may not be tachyonic,
we obtain a total of eighteen 
conditions for the PS-II type models
 with a D6-brane at angles
configuration to be stable.
They are
given in Appendix I.
We don't consider
the scalars from 
the $aa^{\star}$, $dd^{\star}$, $ee^{\star}$ intersections. For these sectors
we will require later that they preserve $N=1$ SUSY. 
As a result all scalars in these sectors may become
massive or receive vevs and becoming eventually massive.

\begin{figure}
\centering
\epsfxsize=6in
\hspace*{0in}\vspace*{.2in}
\epsffile{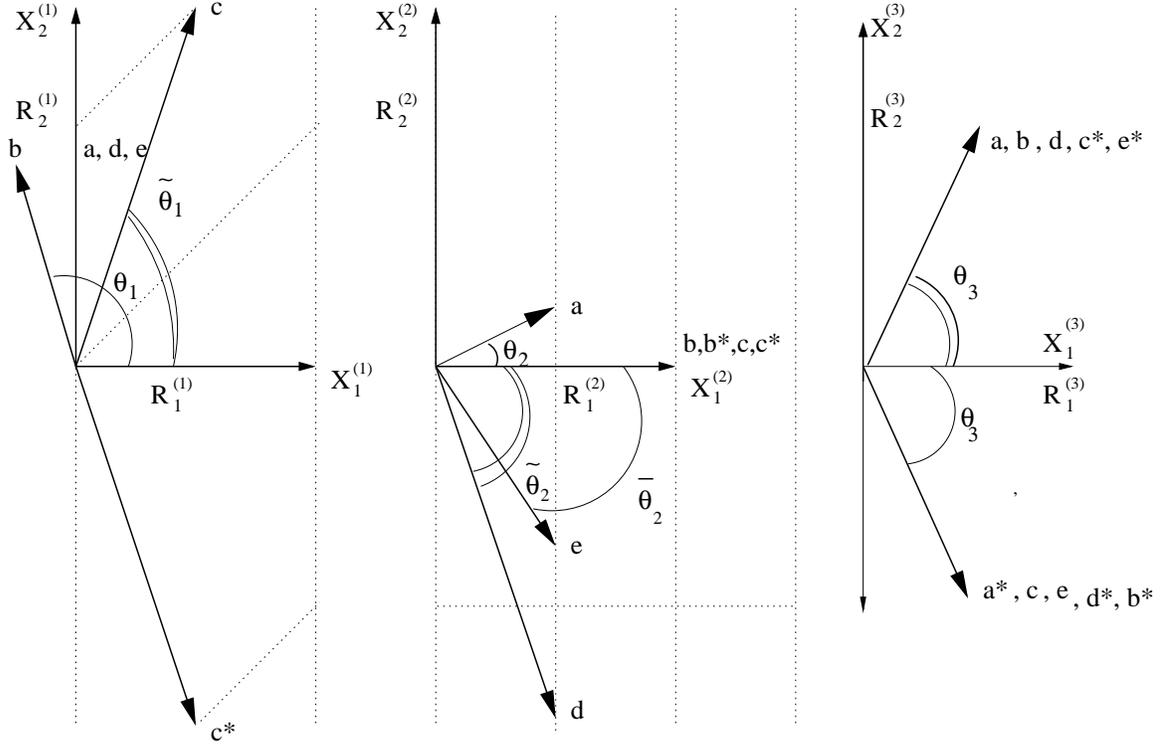}
\caption{\small
Assignment of angles between D6-branes on the
PS-II class of models
based on the initial gauge group $U(4)_C\times {U(2)}_L\times
{U(2)}_R$. The angles between branes are shown on a product of 
$T^2 \times T^2 \times T^2$. We have chosen  $m_b^1, 
m_c^1, n_a^2, n_d^2, n_e^2  >0$,$\epsilon = {\tilde \epsilon}= 1$.
These models
break to low energies to exactly the SM.}
\end{figure}

Lets us now turn our discussion to the Higgs sector of 
PS-II models.
In general there are two different Higgs fields that may
be used to break the PS symmetry.
We remind that they were given in (\ref{useful1}).
The question is if $H_1$, $H_2$ are present in the spectrum of 
PS-II models.
In general, tachyonic scalars stretching between two 
different branes $\tilde a$, 
$\tilde b$, can be used as Higgs scalars as they can become non-tachyonic
by varying the distance between the branes.
Looking at the $I_{a c^{\star}}$ intersection we can confirm that 
the scalar doublets $H^{\pm}$ get localized.
They come from open strings
stretching
between the $U(4)$ $a$-brane and $U(2)_R$ $c^{\star}$-brane.

\begin{table} [htb] \footnotesize
\renewcommand{\arraystretch}{1}
\begin{center}
\begin{tabular}{||c|c||c|c|c|c||}
\hline
\hline
Intersection & PS breaking Higgs & $Q_a$ & $Q_b$ & $Q_c$ & $Q_d$ \\
\hline\hline
$a c^{\star}$ & $H_1$  &
$1$ & $0$ & $1$ & $0$ \\
\hline
$a c^{\star}$  & $H_2$  & $-1$ & $0$ & $-1$ & $0$  \\
\hline
\hline
\end{tabular}
\end{center}
\caption{\small Higgs fields responsible for the breaking of 
$SU(4) \times SU(2)_R$ 
symmetry of the 
$SU(4)_C \times SU(2)_L \times SU(2)_R$ with D6-branes
intersecting at angles. These Higgs are responsible for giving
masses to the right handed
neutrinos in a single family.
\label{Higgs}}
\end{table}

The $H^{\pm}$'s come from the NS sector and
correspond to the states \footnote{a similar set of states was used
in \cite{louis2} to provide the model with electroweak Higgs scalars.}  
{\small \beqa
\begin{array}{cc}
{\rm \bf State} \quad & \quad {\bf Mass^2} \\
(-1+\vartheta_1, \vartheta_2, 0, 0) & \alpha' {\rm (Mass)}^2_{H^{+}} =
  { {Z_3}\over {4\pi ^2}}\ +\ \frac{1}{2}(\vartheta_2 - 
\vartheta_1) \\
(\vartheta_1, -1+ \vartheta_2, 0, 0) & \alpha' {\rm (Mass)}^2_{H^{-}} =
  { {Z_3}\over {4\pi ^2 }}\ +\ \frac{1}{2}(\vartheta_1
  - \vartheta_2) \\
\label{Higgsmasses}
\end{array}
\eeqa}
where $Z_3$ is the distance$^2$ in transverse space along the third torus, 
$\vartheta_1$, $\vartheta_2$ are the (relative)angles 
between the $a$-, $c^{\star}$-branes in the 
first and second complex planes respectively.  
The presence of scalar doublets $H^{\pm}$ can be seen as
coming from the field theory mass matrix

\beq
(H_1^* \ H_2) 
\left(
\bf {M^2}
\right)
\left(
\begin{array}{c}
H_1 \\ H_2^*
\end{array}
\right) + h.c.
\eeq
where
\beqa
{\bf M^2}=M_s^2
\left(
\begin{array}{cc}
Z_3^{(ac^*)}(4 \pi^2)^{-1}&
\frac{1}{2}|\vartheta_1^{(ac^*)}-\vartheta_2^{(ac^*)}|  \\
\frac{1}{2}|\vartheta_1^{(ac^*)}-\vartheta_2^{(ac^*)}| &
Z_3^{(ac^*)}(4 \pi^2)^{-1}\\
\end{array}
\right),
\eeqa
\vspace{1cm}
The fields $H_1$ and $H_2$ are thus defined as
\beq
H^{\pm}={1\over2}(H_1^*\pm H_2) 
\eeq
where their charges are given in table (\ref{Higgs}). 
Hence the effective potential which 
corresponds to the spectrum of the PS symmetry breaking
Higgs scalars is given by
\beqa
V_{Higgs}\ =\ m_H^2 (|H_1|^2+|H_2|^2)\ +\ (m_B^2 H_1H_2\ +\ h.c)
\label{Higgspot}
\eeqa
where
\beqa 
{m_H}^2 \ =\ {{Z_3^{(ac^*)}}\over {4\pi ^2\alpha '}} &;&
m_B^2\ =\ \frac{1}{2\alpha '}|\vartheta_1^{(ac^*)}-
\vartheta_2^{(ac^*)}|
\label{masillas}
\eeqa
The precise values of $m_H^2$, $m_B^2$, for PS-II
classes of
 models are given by
\beqa
 {m_H}^2 \ \stackrel{PS-II}{=}\ {
 {(\xi_a^{\prime}+\xi_c^{\prime})^2}\over {\alpha '}},&&
m_B^2\ \stackrel{PS-II}{=}\
\frac{1}{2\alpha '}|\frac{1}{2}+ {\tilde\theta}_1 -
{\theta}_2|\ ,
\label{value100}
\eeqa
where $\xi_a^{\prime}$($\xi_c^{\prime}$) is the distance between the
orientifold plane
and the $a$($c$) branes and ${\tilde \theta}_1$, 
$ {\theta}_2$ were defined in
(\ref{angPSII}). Thus

\beqa
m_B^2 &\stackrel{PS-II}
{=}&
\frac{1}{2}|
m^2_{\chi_R^2} (t_2) +\ m^2_{\chi_R^2}(t_3)
-\ m^2_{{F}_L} (t_1) -\ m^2_{{F}_L}(t_3) |\nonumber\\
&=&
\frac{1}{2}| m^2_{\chi_R^2} (t_2)
+\ m^2_{\chi_R^2 }(t_3)
-m^2_{{\bar F}_R} (t_1) -\ m^2_{{\bar F}_R}(t_3)|
\nonumber\\
\label{kainour}
\eeqa

For PS-II models the number of Higgs present
is equal to the 
the intersection number product between the $a$-, $c^{\star}$- branes
in the
first and second complex planes, 
\beq
n_{H^{\pm}} \stackrel{PS-I}{=}\
I_{ac^{\star}}\ =\ 3.
\label{inter1}
\eeq
A comment is in order.   
For PS-II models the number of PS Higgs is three.
That means that we have
three
intersections and to each one we have a Higgs particle
which is a linear
combination of the Higgs $H_1$ and $H_2$.

The electroweak symmetry breaking could be delivered through
the bidoublets Higgs present
in the $bc^{\star}$ intersection (seen in table
(\ref{Higgs3}). In principle these 
can be used to give mass ot quarks and leptons.
In the present models their number is
given by
\begin{table} 
\begin{center}
\begin{tabular}{||c|c||c|c|c|c||}
\hline
\hline
Intersection & Higgs & $Q_a$ & $Q_b$ & $Q_c$ & $Q_d$ \\
\hline\hline
$bc^{\star}$  & $h_1 = (1, 2, 2)$  &
$0$ & $1$ & $1$ & $0$ \\
\hline
$bc^{\star}$ & $h_2 = (1, {\bar 2}, {\bar 2})$  & $0$ & $-1$ & $-1$ & $0$  \\
\hline
\hline
\end{tabular}
\end{center}                 
\caption{\small Higgs fields present in the intersection $bc^{\star }$
of the
$SU(4)_C \times SU(2)_L \times SU(2)_R$ classes of models
with D6-branes
intersecting at angles. These Higgs give masses to the quarks and
leptons in a single family and could have been
responsible for
electroweak symmetry breaking if their net number was not
zero.
\label{Higgs3}}
\end{table}
the intersection number of the 
$b$, $c^{\star}$ branes in the first tori
\beq
n_{h_1,\  h_2}^{b c^{\star}}\
\stackrel{PS-II}{=}\ |\epsilon(m_c^1 - m_b^1)|  =\
|\beta^2 (2n_a^2 + n_d^1 -n_e^2)|
\label{nadou1}
\eeq
A comment is in order. Because the number of the
electroweak
bidoublets in the PS-II models depends on the difference
$|m_b^1-m_c^1|$, given the conditions for N=1 SUSY in some 
sectors in the models (see (\ref{modede1}),
(\ref{modede2}) at next section),
we get $n_{h^{\pm}} = 0$ and thus $m_b^1 =\ m_c^1$.
However, this is not a problem for electroweak symmetry breaking as 
(see section 8) a different term is used to provide Dirac 
masses to quarks, leptons and neutrinos. In the present
models is it
important that 
\beq
I_{bc} =\ |m_c^1 + m_b^1| = 2 |m_b^1|
\label{nadou112}
\eeq
may be chosen different from zero.
Thus an alternative set of electroweak Higgs may be
provided from the 
the NS sector where
the lightest scalar states $h^{\pm}$ originate from
open strings stretching between the
$bc$ branes, e.g. named as $h_3$, $h_4$.

\begin{table} 
\begin{center}
\begin{tabular}{||c|c||c|c|c|c||}
\hline
\hline
Intersection & Higgs & $Q_a$ & $Q_b$ & $Q_c$ & $Q_d$ \\
\hline\hline
$bc$  & $h_3 = (1, {\bar 2}, 2)$  &
$0$ & $-1$ & $1$ & $0$ \\
\hline
$bc$ & $h_4 = (1, 2, {\bar 2})$  & $0$ & $1$ & $-1$ & $0$  \\
\hline
\hline
\end{tabular}
\end{center}                 
\caption{\small Higgs fields present in the
intersection $bc$
of the
$SU(4)_C \times SU(2)_L \times SU(2)_R$ classes of models
with D6-branes
intersecting at angles. These Higgs give masses to the
quarks and
leptons in a single family and are
responsible for
electroweak symmetry breaking.
\label{Higgs33}}
\end{table}

{\small \beqa
\begin{array}{cc}
{\rm \bf State} \quad & \quad {\bf Mass^2} \\
(-1+\vartheta_1, 0, \vartheta_3, 0) & \alpha' {\rm (Mass)}^2 =
  \frac{Z_2^{bc}}{4\pi^2}\ +\
  \frac{1}{2}(\vartheta_3 - \vartheta_1)  \\
 (\vartheta_1, 0, -1+\vartheta_3, 0) &  \alpha' {\rm (Mass)}^2 =
  \frac{Z_2^{bc}}{4\pi^2} +\
  \frac{1}{2}(\vartheta_1 -\vartheta_3 )
\label{Higgsacstar}
\end{array}
\eeqa}
where $Z_2^{bc}$ is the relative distance in 
transverse space along the second torus from the orientifold plane, 
$\vartheta_1$, $\vartheta_3$, are the (relative)angle 
between the $b$-, $c$-branes in the 
first and third complex planes.

Hence the presence of scalar doublets $h^{\pm}$
defined as
\beq
h^{\pm}={1\over2}(h_3^*\pm h_4) \   \ .
\eeq
can be seen
as
coming from the field theory mass matrix

\beq
(h_3^* \ h_4) 
\left(
\bf {M^2}
\right)
\left(
\begin{array}{c}
h_3 \\ h_4^*
\end{array}
\right) + h.c.
\eeq
where
\beqa
{\bf M^2}=M_s^2
\left(
\begin{array}{cc}
Z_{23}^{(bc)}(4\pi^2)^{-1}&
\frac{1}{2}|\vartheta_1^{(bc)}-\vartheta_3^{(bc)}|  \\
\frac{1}{2}|\vartheta_1^{(bc)}-\vartheta_3^{(bc)}| &
Z_{23}^{(bc)}(4\pi^2)^{-1}\\
\end{array}
\right),
\eeqa
The effective potential which 
corresponds to the spectrum of electroweak
Higgs $h_3$, $h_4$ may be written as
\beqa
V_{Higgs}^{bc}\ =\ \overline{m}_H^2 (|h_3|^2+|h_4|^2)\ +\ 
(\overline{m}_B^2 h_3 h_4\ +\ h.c)
\label{bcstarpote}
\eeqa
where
\beqa 
\overline{m}_H^2 \ =\ \frac{Z_2^{(bc)}}{4\pi^2\alpha'} \
& ;&
\overline{m}_B^2\ =\ \frac{1}{2\alpha'}|
\vartheta_1^{(bc)} - \vartheta_3^{(bc)}|
\label{bchiggs}
\eeqa
The precise values for PS-II classes of
models $\overline{m}_H^2$, $\overline{m}_B^2$ are
\beqa 
 {\bar m}_{H}^2 \ \stackrel{PS-II}{=}\ { {({\tilde \chi}_b^{(2)}
 +{\tilde \chi}_{c^{\star}}^{(2)} )^2}\over
{\alpha '}}\ ;\
{\bar m}_{B}^2\ \stackrel{PS-II}{=}\ \frac{1}{2\alpha'}
|\theta_1 - \tilde\theta_1 - 2 \theta_3|
\ ;
\label{value1002}
\eeqa
where $\theta_1$,  $\tilde \theta_1$, $\theta_3$ were defined
in (\ref{angPSII}).
Also ${\tilde \chi}_b, {\tilde \chi}_{c^{\star}}$ are the 
distances of the $b$, $c$ branes from the
orientifold plane in the second tori.
The values of the angles $\vartheta_1$,
$\tilde \vartheta_1$, $\vartheta_2$,
can be expressed
in terms of the scalar masses in the various
intersections. We list them for convenience
\beqa
\frac{1}{\pi}\theta_1 &=& \frac{1}{2}|-1 +\
 m^2_{ab^{\star}} (t_2) +\ m^2_{ab^{\star}}(t_3) | \nonumber\\
&=& \frac{1}{2}|1 +\
 m^2_{be} (t_2) +\ m^2_{be}(t_3) | \nonumber\\
&=&  \frac{1}{2}|-1 +\
m^2_{bd^{\star}} (t_2) +\ m^2_{bd^{\star}}(t_3)|
\eeqa
\beqa
\frac{1}{\pi}{\tilde \theta}_1 &=&  \frac{1}{2}|1 -\
m^2_{ac} (t_2) -\ m^2_{ac}(t_3) |
\nonumber\\
&=& \frac{1}{2}|1 -\ m^2_{cd} (t_2) +\ m^2_{cd}(t_3) |
\nonumber\\
&=& \frac{1}{2}|- 1 +\ m^2_{ce^{\star} }
(t_2) +\ m^2_{ce^{\star} }(t_3)|
\label{theta1}
\eeqa
\beqa
\frac{1}{\pi}\theta_2 &=& \frac{1}{2}|
 m^2_{ab^{\star}} (t_1) +\ m^2_{ab^{\star}}(t_3) | \nonumber\\
&=& \frac{1}{2}|
 m^2_{ac} (t_1) +\ m^2_{ac}(t_3) |
\eeqa

\beqa
\frac{1}{\pi}\theta_3 &=& \frac{1}{4}|
 m^2_{F_L} (t_1) +\ m^2_{F_L}(t_2) | = \frac{1}{4}|
 m^2_{{\bar F}_R} (t_1) +\ m^2_{{\bar F}_R}(t_2) | \nonumber\\
&=& \frac{1}{4}|
 m^2_{\chi_L^1} (t_1) +\ m^2_{\chi_L^1}(t_2) | = \frac{1}{4}|
 m^2_{\chi_R^1} (t_1) +\ m^2_{ \chi_R^1}(t_2) | \nonumber\\
&=& \frac{1}{4}|
 m^2_{\chi_L^2} (t_1) +\ m^2_{\chi_L^2}(t_2) | = \frac{1}{4}|
 m^2_{\chi_R^2} (t_1) +\ m^2_{ \chi_R^2}(t_2) |
\eeqa

\section{Singlet scalar generation - $N=1$ SUSY on Intersections}

In this section, we intend to demand that
certain open string sectors
respect $N=1$ supersymmetry.
In particular we will focus in showing how
by we can create sectors which preserve N=1
supersymmetry in non-supersymmetric intersecting
D6-brane models.
This is most interesting
for the good phenomenology of the models, as
SUSY sectors guarantee the generation of singlets.
The singlet scalars
will be necessary for giving masses to the extra $U(1)$
gauge bosons which they don't have any couplings to
the RR fields and also realizing a majotana mass term
for the right handed neutrinos \footnote{Withour N=1 SUSY this coupling would 
have been absent and thus the models useless for good phenomenology as in the 
base case the
right handed neutrinos would survive massless to low energy. See 
also \cite{kokos1} and section (8.2).}
We note that the spectrum
of PS-II classes of models described in
table (\ref{spectrum8}) is massless at this point
Thus supersymmetry will create singlet scalars
which receive vevs and generate masses for the 
otherwise massless
fermions $\chi_L^1$, $\chi_L^2$, $\chi_R^1$, $\chi_R^2$,
$\omega_L$, $y_R$, $s_R^1$, $s_R^2$. For the status of
vevs in the context of intersecting branes see a
relevant comment on
the concluding section.

Before presenting the analysis, let us
note that a Majorana mass term for right neutrinos
appears only once we impose $N=1$ SUSY on
an intersection.
That will have as an effect the appearance of the
massless scalar
superpartners
of the ${\bar F}_R$ fermions, the ${\bar F}_R^H$'s,
thus allowing a dimension 5
Majorana mass term for $\nu_R$,
$F_R F_R {\bar F}_R^H {\bar F}_R^H$.
The see-saw mechanism for the Majorana
neutrinos will be discussed in section 8.

\subsection{\em PS-II models with N=1 SUSY}

In this part we will show that model dependent
conditions,
obtained by demanding that the extra U(1)'s do not have
non-zero couplings to the RR fields, are necessary
conditions in order to have scalar singlet generation
that could effectively break the extra U(1)'s.
These conditions will be alternatively obtained by demanding that certain
string sectors respect $N=1$ supersymmetry.

In general, for $N=1$ supersymmetry to be
preserved at some
intersection
between two branes {\em L}, {\em M}, we need to
satisfy
$\pm \vartheta_{ab}^1 \pm \vartheta_{ab}^2 \pm
\vartheta_{ab}^3$ for
some choice of signs, where
$\vartheta_{\alpha \beta}^i$, $i=1,2,3$
are the relative angles of the
branes {\em L}, {\em M} across the three 2-tori.
The latter rule will be our main tool in getting
N=1 SUSY on intersections.

\begin{itemize}

\item {\em The $ac$ sector respects ${\cal N} =1$
supersymmetry.}

The condition for $N=1$ SUSY on the $ac$-sector
is \footnote{We have chosen $m_c^1 < 0$.}:
\beq
\pm (\frac{\pi}{2} +\ {\tilde \vartheta}_1) \ \pm  {\vartheta}_2 \
\pm  2\vartheta_3
\ = 0,
\label{condo1}
\eeq       
This condition can be solved by choosing :
\beq
ac \rightarrow (\frac{\pi}{2} + \ {\tilde \vartheta}_1) \
+ \vartheta_2 \
- 2\vartheta_3 \ = 0,
\label{condo10}
\eeq       
and thus may be solved by the choice \footnote{We have set
$U^{(i)}= \frac{R_2^{(i)}}{R_1^{(i)}}$, $i=1, 2,3$}
\beq
-{\tilde \vartheta}_1 \ =
\vartheta_2 \ = \vartheta_3 \ = \frac{\pi}{4},
\label{solver1}
\eeq
effectively giving us
\beq
-  \frac{1}{\epsilon m_c^1 \ U^{(1)}} = \
\frac{(\epsilon {\tilde \epsilon}) n_a^2}{3  \b_2 U^{(2)}}
= \
  \frac{2 {\tilde \epsilon}}{U^{(3)}} = \ \frac{\pi}{4}.
\label{condo3}
\eeq
By imposing $N=1$ SUSY on an intersection ac the 
massless scalar
superpartner of ${\bar F}_R$ appears, the ${\bar F}_R^B$.
Note that in (\ref{condo3}) the imposition of N=1 SUSY
connects the complex structure moduli $U^i$ in the different tori
and thus reduces the moduli degeneracy of the theory.

\end{itemize}

\begin{itemize}

\item {\em The $dd^{\star}$ sector preserves
${\cal N} = 1$ supersymmetry}

As we noted in the appendix the presence of N=1
supersymmetry in the sectors $dd^{\star}$,
$ee^{\star}$ is equivalent to the absence of tachyons in
those sectors.

The general form of the ${\cal N} = 1$
supersymmetry condition on this sector
is
\beq
\pm \pi \pm 2 {\tilde \vartheta}_2 \ \pm  2
{\vartheta}_3 \
\ = 0,
\label{condo1oup1}
\eeq       
which may be solved by the choice
\beq
 \pi \ + 2 {\tilde \vartheta}_2 \ -  2
{\vartheta}_3 \
\ = 0,
\label{esosi1}
\eeq       
Hence
\beq
-{\tilde \vartheta}_2 \ = \vartheta_3 \ = \frac{\pi}{4},
\label{solv1sol23}
\eeq
that is
\beq
-\frac{\epsilon {\tilde \epsilon}n_d^2}{4 \beta_2} U^{(2)} = \
  \frac{2}{{\tilde \epsilon}}U^{(3)} = \ \frac{\pi}{4}.
\label{laop13}
\eeq

\end{itemize}

\begin{itemize}

\item {\em The $ee^{\star}$ sector preserves
${\cal N} = 1$ supersymmetry}

The general form of the ${\cal N} = 1$
supersymmetry condition on this sector
is
\beq
\pm \pi \pm\ 2 {\bar \vartheta}_2 \ \pm  2
{\vartheta}_3 \
\ = 0,
\label{oppprae1}
\eeq       
which we may recast in the form
\beq
- \pi +\ 2 {\bar \vartheta}_2 \ + 2
{\vartheta}_3 \
\ = 0,
\label{allo2}
\eeq       
be solved by the choice
\beq
{\bar  \vartheta}_2 \ = \vartheta_3 \ = \frac{\pi}{4},
\label{condo23}
\eeq
that is
\beq
\frac{ (\epsilon {\tilde \epsilon}) n_e^2}{2 \beta_2 U^{(2)}} = \
  \frac{2}{{\tilde \epsilon}}U^{(3)} = \ \frac{\pi}{4}.
\label{laopzop13}
\eeq

From (\ref{laopzop13}), (\ref{laop13}), (\ref{condo3}), 
 we derive the conditions
\beq 
2 n_a^2 = 3 n_e^2,
\label{modede1}
\eeq
\beq 
 n_d^2 = -2  n_e^2 \ .
\label{modede2}
\eeq
\end{itemize}

An important comment is in order.  The presence
of N=1 supersymmetry in $dd^{\star}$, $ee^{\star}$
sectors signals the presence of the scalar
superpartners of $s_R^1$, $s_R^2$, namely the 
$s_B^1$, $s_B^2$ respectively. The latter scalars may
receive vevs.
Thus imposing N=1 SUSY in leptonic sectors guarantee
the presence of gauge singlets in the models.

Also what is is evident by looking at conditions
(\ref{ena11}), (\ref{modede1}), (\ref{modede2}) 
is that the
conditions of orthogonality for the extra $U(1)$'s
to survive massless the generalized Green-Schwarz
mechanism is equivalent to the conditions for N=1
supersymmetry in the leptonic sectors $dd^{\star}$,
$ee^{\star}$.  
The latter condition is equivalent to the
absence of tachyons in the sectors $dd^{\star}$,
$ee^{\star}$ as someone might check.

A numerical set of wrapping
numbers consistent with the RR tadpole
constraints (\ref{ena11}), (\ref{modede1}) and 
(\ref{modede2}) is
\beq
\epsilon =\ {\tilde \epsilon} =\ 1,\ n_a^2=15,
\;m_b^1=1,\ m_c^1= 1, \
n_d^2=-20,\ n_e^2=10,
\;\b_1=1, \  \b_2=1/2.
\label{numero11}
\eeq
The latter can be satisfied with the addition of four
extra U(1)
D6-branes.
\begin{table}[htb]\footnotesize
\renewcommand{\arraystretch}{2}
\begin{center}
\begin{tabular}{||c||c|c|c||}
\hline
\hline
$N_i$ & $(n_i^1, m_i^1)$ & $(n_i^2, m_i^2)$ & $(n_i^3, m_i^3)$\\
\hline\hline
 $N_a=4$ & $(0, 1)$  &
$(15, 3/2)$ & $(1, 1/2)$  \\
\hline
$N_b=2$  & $(-1, 1 )$ & $(2, 0)$ &
$(1, 1/2)$ \\
\hline
$N_c=2$ & $(1, 1 )$ &   $(2, 0)$  & 
$(1, -1/2)$ \\    
\hline
$N_d=1$ & $(0, 1)$ &  $(-20, -2)$
  & $(2, 1)$  \\   
\hline
$N_e=1$ & $(0, 1)$ &  $(10, -1)$
  & $(2, -1)$  \\   
\hline
\end{tabular}
\end{center}
\caption{\small Wrapping number set consistent
with the tadpole constraint (\ref{ena11}) and
the $N=1$ SUSY preserving conditions
(\ref{modede1}), (\ref{modede2}).
The SUSY conditions can be shown to be equivalent
to the model dependent conditions for $U(1)$'s surviving
massless the generalized Green-Schwarz mechanism
(see section 5).
 We have not included
the extra U(1) branes.
\label{teley}}
\end{table}

We note that the constraint (\ref{ena11}) is independent
from
the number of extra U(1)'s added to satisfy the RR tadpole
conditions. Thus we may safely
choose $\beta =1$, $\beta=1/2$ that is $N_h = 4$ in
table (\ref{teley}) positioned at (\ref{2ndform}).
We note that the entries of the d-brane in the second
tori is such that it appears to be corresponding to
a U(2) brane wrapping once around
or a U(1) brane wrapping twice 
along the one-cycle on the 2nd tori. However, as has been
noted before \footnote{see the first reference
of \cite{kokos1}.}
this is not a problem since a multiwrapping of this form
can be absorbed in a $U(1)_d$ field redefinition in e.g.
the anomaly free $U(1)^{(4)}$, that survives massless the
generalized Green-Schwarz mechanism in section 5.

\subsection{\em Gauge singlet generation from the extra
U(1) branes}

In this section, we will present an alternative
mechanism for generating singlet scalars.
We had already seen that in leptonic sectors involving
U(1) branes, e.g. $dd^{\star}$, $ee^{\star}$, brane
imposing
N=1 SUSY creates singlet scalars. This is reflected
in the fact that in U(1) $j$-branes, sectors in the form
$jj^{\star}$ had localized in their intersection
gauge singlet fermions. Thus imposing N=1 SUSY on those
sectors help us to get rid of these masslesss fermions,
by making them
massive through their couplings to their superpartner
gauge singlet scalars.

What we will become clear in this sector is that the
presence of supersymmetry in particular sectors involving
the extra branes creates singlet scalars that 
provide the couplings that make massive
some non-SM fermions.

In order to show the creation of gauge singlets
from sectors involving extra branes we will make our
points
by using only one of the extra $N_h$ $U(1)$ branes, e.g. the
$N_{h_1}$ one. The following
discussion can be identically repeated for the other
extra branes.

Thus due to the non-zero intersection numbers
of the
$N_{h_1}$ U(1) brane with $a$,$d$ branes the following
sectors are
present : $ah$, $ah^{\star}$, $dh$, $dh^{\star}$.

\begin{itemize}

\item $ah$-{\em sector}

Because $I_{ah} = -\frac{3}{\beta_1}$
we have present $|I_{ah}|$ massless fermions
$\kappa_1^f$ in the representations
\beq
\kappa_1^f \rightarrow \ ({\bar 4}, 1, 1)_{(-1, 0, 0, 0, 0; 1)}
\label{ah}
\eeq
where the subscript last entry denotes the U(1) charge
of the `sixth' U(1) extra brane \footnote{We don't exhibit
the beyond the sixth entry of the rest of the extra
branes as they are obviously zero for the present discusion.}.

\item $ah^{\star}$-{\em sector}

Because $I_{ah^{\star}}= -\frac{3}{\beta_1} < 0$,
there are present
$|I_{ah^{\star}}|$
fermions $\kappa_2^f$ appearing as a linear
combination of the representations
\beq
\kappa_2^f \rightarrow \ ({\bar  4}, 1, 1 )_{(-1, 0, 0, 0, 0; -1)}
\l{dh2}
\eeq

\item $dh$-{\em sector}

Because $I_{dh} = \frac{8}{\beta_1}$, there are present
$|I_{dh}|$ fermions $\kappa_3^f$ transforming in the representations
\beq
\kappa_3^f \rightarrow \ (1, 1, 1)_{(0, 0, 0, 1, 0; -1)}
\label{reps1}
\eeq

We further require that this sector respects
$N=1$ supersymmetry.
In this case we have also present the massless scalar
fields $\kappa_3^B$,
\beq
\kappa_3^B \rightarrow \ (1, 1, 1 )_{(0, 0, 0, 1, 0; -1)},
\l{aradh5}
\eeq
The latter scalars receive a vev which we assume to be
of order of the string scale.

 The condition for
$N=1$ supersymmetry in this sector is exactly
\beq
-\frac{\pi}{2} + {\tilde \vartheta}_2
+ (\vartheta_3) = 0
\label{exactamente1}
\eeq
which is satisfied when ${\tilde \vartheta}_2$,
${\vartheta}_3$ take the value $\pi/4$ in consistency
with (\ref{solv1sol23}) and subsequently (\ref{allo2}).

\item $dh^{\star}$-{\em sector}

Because $I_{dh^{\star}} =  \frac{8}{\beta_1} \neq 0$,
there are present
$|I_{ah^{\star}}|$
fermions $\kappa_4^f$ in the representations
\beq
\kappa_4^f \rightarrow \ ( 1, 1, 1 )_{(0, 0, 0, 1, 0; 1)}
\l{dh4}
\eeq

The condition that this sector respects N=1 SUSY is
equivalent to the one is the $dh$-sector.

\item $eh$-{\em sector}

Because $I_{eh} = -\frac{4}{\beta_1}$, there are
present
$|I_{eh}|$ fermions $\kappa_5^f$ transforming in the
representations
\beq
\kappa_5^f \rightarrow \ (1, 1, 1)_{(0, 0, 0, 0, -1; 1)}
\label{erepsec1}
\eeq

Also we require that this sector
preserves N=1 SUSY.
Because of N=1 SUSY and $I_{eh} = -\frac{4}{\beta_1}$,
there are
present
$|I_{eh}|$ bosons $\kappa_5^B$ transforming in the
representations
\beq
\kappa_5^B \rightarrow \ (1, 1, 1)_{(0, 0, 0, 0, -1; 1)}
\label{boso1}
\eeq
The condition for N=1 SUSY is
\beq
\pm \frac{\pi}{2} \pm {\bar \vartheta}_2 \pm \vartheta_3 =\ 0
\label{halfcond1}
\eeq
which is exactly `half' of the supersymmetry condition
(\ref{oppprae1}). When it is rearranged into the form
\beq
\frac{\pi}{2} + {\bar \vartheta}_2 - \vartheta_3 = \ 0, 
\label{reaa2}
\eeq
it is solved by the choice (\ref{condo23}).
\newline
Summarizing we have found that the conditions
(\ref{modede1}, (\ref{modede2})
derived as the model dependent conditions
of the U(1)'s that survive the generalized Green-Schwarz
mechanism, are equivalent : \newline
$\bullet$ to have the leptonic branes, d, e, preserve N=1
SUSY on the sectors $dd^{\star}$, $ee^{\star}$.
\newline
$\bullet$ to have the sectors made of a mixture of the
extra and leptonic branes preserve N=1 SUSY. The presence of 
these conditions is independent from the number of
extra U(1) branes present.

\item $eh^{\star}$-{\em sector}
In this sector,  $I_{eh^{\star}} = -\frac{4}{\beta_1}$.
 Thus there are
present
$|I_{eh^{\star}}|$ fermions $\kappa_6^f$ transforming in
the
representations
\beq
\kappa_6^f \rightarrow \ (1, 1, 1)_{(0, 0, 0, 0, 0, -1; -1)}
\label{erepsec11}
\eeq
The condition for N=1 SUSY to be preserved by this section
is exactly (\ref{reaa2}). Thus we have present
$|I_{eh^{\star}}|$ bosons $\kappa_6^B$ transforming
in the 
representations
\beq
\kappa_6^B \rightarrow \ (1, 1, 1)_{(0, 0, 0, 0, -1; -1)}
\label{boso2}
\eeq

\end{itemize}

We will now show that all fermions, appearing from
the non-zero intersections of the extra
brane $U(N_{h_1})$
with the branes $a$, $d$,  $e$, receive string scale mass and disappear
from the low energy spectrum (see also a related discussion
in the concluding section).

\begin{itemize}

\item The mass term for the $\kappa_1^f$ fermion reads:
\beqa
(4, 1, 1 )_{(1, 0, 0, 0, 0; -1)} \
(4, 1, 1 )_{(1, 0, 0, 0,tilde  0; -1)}  \
\langle({\bar 4}, 1, 2)_{(-1, 0, 1, 0, 0, ;0)} \rangle \nonumber\\
\times \langle ({\bar 4}, 1, {\bar 2})_{(-1, 0, -1, 0, 0; 0)} \rangle
\langle (1, 1, 1)_{(0, 0, 0, 0, 1; 1)}\rangle \
\langle (1, 1, 1)_{(0, 0, 0, 0, -1; 1)}\rangle
\label{eksiso1}
\eeqa
or
\beq
{\bar \kappa}_1^f \ {\bar \kappa}_1^f \
\langle H_2 \rangle \
\langle {\bar F}_R^H \rangle \
\langle {\bar \kappa}^6_B \rangle \
\langle {\kappa}^5_B \rangle
\sim    {\bar \kappa}_1^f \ {\bar \kappa}_1^f \    M_s
\eeq

\item The mass term for the $\kappa_2^f$ fermion reads:

\beqa
(4, 1, 1 )_{(1, 0, 0, 0, 0; 1)} \
(4, 1, 1 )_{(1, 0, 0, 0, 0; 1)}  \
\langle({\bar 4}, 1, 2)_{(-1, 0, 1, 0, 0, ;0)} \rangle \nonumber\\
\times \langle ({\bar 4}, 1, {\bar 2})_{(-1, 0, -1, 0, 0; 0)} \rangle
\langle (1, 1, 1)_{(0, 0, 0, 0, -1; -1)}\rangle \
\langle (1, 1, 1)_{(0, 0, 0, 0, 1; -1)}\rangle
\label{eksiso2}
\eeqa
or
\beq
{\bar \kappa}_2^f \ {\bar \kappa}_2^f \
\langle H_2 \rangle \
\langle {\bar F}_R^H \rangle \
\langle {\bar \kappa}^5_B \rangle \
\langle {\kappa}^6_B \rangle
\sim    {\bar \kappa}_2^f \ {\bar \kappa}_2^f \    M_s
\eeq

\item The mass term for the $\kappa_3^f$ fermion reads:

\beqa
(1, 1, 1 )_{(0, 0, 0, -1, 0; 1)} \
(1, 1, 1 )_{(-1, 0, 0, -1, 0;  1)} \
\langle (1, 1, 1)_{(0, 0, 0, 1, 0; -1)} \rangle  \
\langle (1, 1, 1)_{(0, 0, 0, 1, 0; -1)} \rangle
\label{eksiso4}
\eeqa
or
\beq
{\bar \kappa}_3^f \ {\bar \kappa}_3^f \
\langle{ \kappa}_3^B \rangle \
\langle {\kappa}_3^B \rangle  \sim \ M_s \ {\bar \kappa}_3^f \ 
{\bar \kappa}_3^f
 \eeq

\item The mass term for the $\kappa_4^f$ fermion reads:

\beqa
(1, 1, 1 )_{(0, 0, 0, -1, 0; -1)} \
(1, 1, 1 )_{(0, 0, 0, -1, 0;  -1)} \
\langle (1, 1, 1)_{(0, 0, 0, 1, 0; 1)} \rangle  \
\langle (1, 1, 1)_{(0, 0, 0, 1, 0; 1)} \rangle
\label{eksiso44}
\eeqa
or
\beq
{\bar \kappa}_4^f \ {\bar \kappa}_4^f \
\langle{ \kappa}_4^B \rangle \
\langle {\kappa}_4^B \rangle  \sim \ M_s \ {\bar \kappa}_4^f \ 
{\bar \kappa}_4^f
 \eeq

\item The mass term for the $\kappa_5^f$ fermion reads:

\beqa
(1, 1, 1 )_{(0, 0, 0, 0, 1; -1)} \
(1, 1, 1 )_{(0, 0, 0, 0, 1; -1)} \
\langle (1, 1, 1)_{(0, 0, 0, -1, 0; 1)} \rangle  \
\langle (1, 1, 1)_{(0, 0, 0, -1, 0; 1)} \rangle
\label{eksiso5}
\eeqa
or
\beq
{\bar \kappa}_5^f \ {\bar \kappa}_5^f \
\langle{ \kappa}_5^B \rangle \
\langle {\kappa}_5^B \rangle  \sim \ M_s \ {\bar \kappa}_5^f \ 
{\bar \kappa}_5^f
 \eeq

\item The mass term for the $\kappa_6^f$ fermion reads:

\beqa
(1, 1, 1 )_{(0, 0, 0, 0, 1; 1)} \
(1, 1, 1 )_{(0, 0, 0, 0, 1; 1)} \
\langle (1, 1, 1)_{(0, 0, 0, -1, 0; -1)} \rangle  \
\langle (1, 1, 1)_{(0, 0, 0, -1, 0; -1)} \rangle
\label{eksiso6}
\eeqa
or
\beq
{\bar \kappa}_6^f \ {\bar \kappa}_6^f \
\langle{ \kappa}_6^B \rangle \
\langle {\kappa}_6^B \rangle  \sim \ M_s \ {\bar \kappa}_6^f \ 
{\bar \kappa}_6^f
 \eeq

\end{itemize}

\subsection{\em Breaking the anomaly free massless U(1)'s}

As in the standard version of a left-right
Pati-Salam $SU(4) \times SU(2)_L \times SU(2)_R$
model, if the
neutral component of $H_1$ (resp. $H_2$), $\nu_H$,
acquires a vev, e.g. $\langle \nu^H \rangle$, then
the initial gauge symmetry,
$SU(4) \times SU(2)_L \times SU(2)_R
\times U(1)_a \times U(1)_b \times U(1)_c \times U(1)_d$,
  can break to the
standard model gauge group
$SU(3) \t U(2) \t U(1)_Y$ augmented by the extra,
non-anomalous,
$U(1)$'s,  
$Q^{(4)}$, $Q^{(5)}$, $Q^{(6)}$, $Q^{(7)}$.
Note that we have considered for simplicity the possibility
of two extra U(1) branes. For convenience the following
discussion will focus on one of the extra branes, as
identical results hold for the other U(1) brane.

The extra $U(1)$'s may be broken by
appropriate Higgsing.
In the PS-A, PS-I  models, by imposing SUSY on the
sectors
$dd^{\star}$, $dh$, $dh^{\star}$ we made it possible
to generate
the appearance of the scalar superpartners 
of the fermions appearing on the respecting intersections.
In the present models the sectors by preserving
$N=1$ SUSY on sectors
$dd^{\star}$, $ee^{\star}$, $dh$, $dh^{\star}$
we have available the singlets
responsible for breaking the U(1)'s that survive
massless the
Green-Schwarz mechanism.
Thus, looking at (\ref{PSIIab1}), $U(1)^{(4)}$ may
break if $s_B^1$ gets a
vev,  $U(1)^{(7)}$ may break if 
 $s_B^2$ gets a vev,
 $U(1)^{(5)}$ may break if $\kappa_3^B$ (or one of the
  $\kappa_4^B$, $\kappa_5^B$,  $\kappa_6^B$) receives a
 vev. Also the extra $U(1)^{(6)}$ brane
 is treated in the same way as $U(1)^{(5)}$.

Note that in this case the extra non-anomalous $U(1)$'s 
have some important phenomenological properties.
In particular they do not charge the PS symmetry
breaking  Higgs
 scalars $H_1$, $H_2$ thus avoiding the appearance of
 axions.

We note that up to this point the only issue remaining
is how we can give non-zero masses to all exotic
fermions of
table (\ref{spectrum8}) beyond those that accommodate
the quarks and leptons of the SM.

\section{Geometrical Yukawa couplings and lepton masses}

In this section, we will examine the mechanism
of generating neutrino masses in the
$SU(4) \times S(2)_L \times SU(2)_R$ classes of PS-II
GUTS. Also we examine some aspects of the geometry
of the Yukawa couplings. Particular emphasis is given
to the exhibition of the couplings giving masses
to all the fermions appearing in table 1, beyond those
making the quarks and lepton structure.

\subsection{\em Yukawa couplings}

Proton decay is one of the most important problems 
of grand unifies theories. In the standard versions of left-right 
symmetric PS models this problem  
is avoided
as B-L is a gauged symmetry but the problem persists in 
baryon number violating operators of sixth order,
contributing to proton
decay. In the PS-I models proton decay is absent as baryon
number
 survives as a
global symmetry to low energies.
That provides for an explanation for the origin of proton stability
in general brane-world scenarios.
Clearly $Q_a = 3 B + L$ and the baryon B
is given by
\beqa
B = \frac{Q_a + Q_{B-L}}{4}.&  
\label{ba1}
\eeqa

In intersecting brane worlds the usual tree level
SM fermion mass generating trilinear Yukawa couplings
between
the fermion states
$F_L^i$, ${\bar F}_R^j$ and the Higgs fields $H^k$ arise
from the
stretching of the worldsheet between the three
D6-branes which cross
at those intersections. In the Pati-Salam GUTS we examine, 
the trilinear Yukawa
is
\beq
Y^{ijk} F_L^i {\bar F}_R^j h^k
\label{roof1}
\eeq
Its general form for a six dimensional torus
is in the
leading order \cite{luis1},
\beq
Y^{ijk}=e^{- {\tilde A}_{ijk}},
\label{yuk1}
\eeq
where ${\tilde A}_{ijk}$ is the worldsheet area
connecting the three vertices.
The areas of each of the two dimensional torus
involved in
this interaction is typically of order one in string units.
In \cite{kokos1} we have
assumed
that the areas of the second and third tori are close to
zero.
In this case, the area of the full Yukawa coupling (\ref{yuk1})
was given in the form 
\beq
Y^{ijk}= e^{-\frac{R_1 R_2}{a^{\prime}} A_{ijk}},
\label{yuk12}
\eeq
where $R_1$, $R_2$ the radii and 
$ A_{ijk}$ the area of the two dimensional tori in the first
complex plane.
Here we exhibit the leading worldsheet correction coming
from the first tori, as it holds for any PS GUT model
constructed as a deformation of the quark and lepton intersection
numbers, e.g. PS-A, PS-I classes in \cite{kokos1} and
the present PS-II classes of models.

Let us analyze a bit further the relation (\ref{yuk12}).
The area of the interaction (\ref{roof1})
in the graphic representation seen in figure 2
is depicted, in the first tori, by the triangle AABBCC, with 
sides a, b, c named
as the branes lying on them \footnote{
This area in simple Euclidean geometry terms is given by
\beq
A = \sqrt{(a-b)^2(a-c)^2 - ((a-b) \cdot (a-c))^2}
\label{eukli}
\eeq}.
A comment is in order at this point.
The classes of GUT models we have been considering
recently \cite{kokos1} and at the present work, have as
 their low energy theory in
energies of order $M_z$ the Standard model.
Their common characteristic is that they represent
{\em deformations} around the basic intersection
structure of the Quark and Lepton structure,
\beq
I_{ab} = 3, \  I_{ac^{\star}} = -3 \ .
\label{deforme}
\eeq
Thus they all share the same intersection numbers along the
`baryonic' $a$ and the left and right `weak'  $b$ and $c$,  D6 branes.
As we will show the area of the trilinear Yukawa
couplings can be reexpressed in a simple form
in terms of
only intersection numbers of the $b$, $c$ branes along which
the bidoublets $h_1$, $h_2$, $h_3$, $h_4$ responsible in general for electroweak symmetry
breaking, are localized.
 This provides us with a
quantitative relation that may be useful is showing the
hierarchies
among neutrino masses in the general left-right
PS GUT models. Assuming that the triagle areas in the 2nd , 3rd tori 
ar close to zero, for the interaction (\ref{roof1}) the 
worldsheet areas 
for (\ref{roof1}) are given
by
\beqa
|A^{(T_2^{(1)})}| =
|m_b^1 -  m_c^1|,\nonumber\\
|A^{(T_2^{(2)})}| = |A^{(T_2^{(3)})}| = 0,
\label{triti}
\eeqa
The universal relations (\ref{triti}) describe the
triangle area
for all classes of models based on the present PS-II
classes of 
GUTS and the PS-A, PS-I of \cite{kokos1}.

\begin{figure}
\centering
\epsfxsize=4in
\hspace*{0in}\vspace*{.2in}
\epsffile{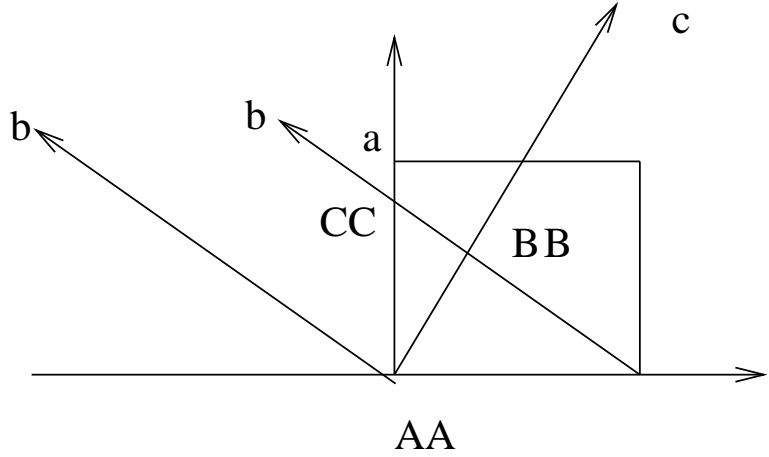}
\caption{\small
Assignment of branes a, b, c 
based on the initial gauge group $U(4)_C\times {U(2)}_L\times
{U(2)}_R$. The angles between branes shown are localized on the 
first tori.·} 
\end{figure}

Hence we have found that the worldsheet area
entering the trilinear Yukuwa couplings is parametrized
in terms
of the parameters describing the RR tadpole conditions.
In the leading
worldsheet instanton expansion the trilinear Yukawa
couplings may be given by
\beq
Y^{ijk} \ F_L \ {\bar F}_R \  h^k \sim \ e^{-\frac{R_1^{(1)}
R_2^{(1)}}{2 a^{\prime}} |m_b^1 - m_c^1|}
\ F_L \ {\bar F}_R \ h^k
\label{firstlaedin}
\eeq
In the present models $m_b^1-m_c^1 = 0$.

\subsection{\em Neutrino masses}

In the present class of GUTS the electroweak
bidoublets (\ref{roof1}) are absent at tree level.
Fortunately for us, there is another coupling, which
is non-renormalizeble, of the same order as the tree
level one.
It is given by      
\beq
F_L \ {\bar F}_R \
\langle h_3 \rangle \
\langle F_R^B \rangle \langle H_2 \rangle \ \sim \
\upsilon  \
F_L \ {\bar F}_R
\label{blue1}
\eeq

For a dimension five interaction term, like those involved in
the
 Majorana mass term for the right handed neutrinos
the interaction term is in the form
\beq
Y^{lmni}= e^{- {\tilde A}_{lmni}},
\label{yuk14}
\eeq
where ${\tilde A}_{lmni}$ the worldsheet area connecting the four interaction
vertices. Assuming that the areas of the second and third  `tetragonal' are
close to zero, the four term coupling can be approximated as
\beq
Y^{ijk}= e^{-\frac{R_1 R_2}{a^{\prime}} A_{lmni}},
\label{yuk15}
\eeq
where the area of the $A_{lmni}$ may be of order one in string units.
\newline
Thus the full Yukawa interaction for the chiral spectrum
of the PS-II models
\beq
\lambda_1 F_L \ {\bar F}_R \ \langle h_3 \rangle \
\langle F_R^B \rangle \langle H_2 \rangle \
+\ \lambda_2 \frac{F_R {F}_R \langle {\bar F}_R^H \rangle
\langle {\bar F}_R^H \rangle}{M_s},
\label{era1} 
\eeq
where
\beqa
\lambda_1 \equiv e^{-\frac{R_1 R_2 A_1}{\alpha^{\prime}}},&
\lambda_2 \equiv e^{-\frac{R_1 R_2 A_2}{\alpha^{\prime}}}.
\label{aswq123}
\eeqa
and the Majorana coupling involves the massless
scalar \footnote{Of order
of the string scale.}  
superpartners ${\bar F}_R^H$ 
of the antiparticles ${\bar F}_R$.
This coupling is unconventional, in the
sense that the ${\bar F}_R^H$ is generated by
imposing SUSY on a
sector of a non-SUSY model.
 We note the
 presence of $N=1$ SUSY at the sector $ac$.
 As can be seen by comparison
 with (\ref{na368}) 
 the ${\bar F}_R^H$ has a neutral direction that
 receives the vev $<H>$.
There is no restriction on the vev of
$F_R^H$ from first
principles and its vev can be anywhere between
the scale of electroweak symmetry breaking and $M_s$.
\newline
The Yukawa term
\beqa
F_L \ {\bar F}_R \
\langle h_3 \rangle \
\langle F_R^B \rangle \ \langle H_2 \rangle \ \sim \
\upsilon  \
F_L \ {\bar F}_R
\label{yukbre12}
\eeqa
is responsible for the electroweak 
symmetry breaking. This term generates Dirac
masses to up quarks and
neutrinos.
Thus, we                  
get
\beq
\lambda_1 F_L \ {\bar F}_R \
\langle h_3 \rangle \ \
\langle F_R^B \rangle \ \langle H_2 \rangle \
  \rightarrow (\lambda_1 \  \upsilon)
(u_i u_j^c + \nu
_i N_j^c) + (\lambda_1 \  {\tilde \upsilon})
\cdot (d_i d_j^c + e_i e_j^c),  
\label{era2}
\eeq
where we have assumed that 
\beq
\langle h_3 \rangle \
\langle F_R^B \ \rangle \langle H_2 \rangle \
= \left(
\begin{array}{cc}
\upsilon  & 0 \\
0 & {\bar \upsilon}
\label{era41}
\end{array}
\right)
\label{finalhiggs}
\eeq
We observe that the model gives non-zero tree level
masses
to the fields present. 
These mass relations may be retained at tree level only,
since as the model
has a non-supersymmetric fermion spectrum, it will
receive higher order corrections.  
It is interesting that from (\ref{finalhiggs})
we derive the GUT relation
\cite{ellis}
\beq
m_d =\ m_e \ .
\label{gutscale}
\eeq
as well the unwanted
 \beq
m_u =\ m_{N^c \nu} \ .
\label{gutscale1}
\eeq 

In the case of neutrino masses, 
the  ``unwanted'' (\ref{gutscale1}), associated
to the $\nu - N^c$
mixing,
is modified due to the presence of the Majorana term
in (\ref{era1})
leading to the see-saw mixing type neutrino mass
matrix 
\beqa
\left(
\begin{array}{cc}
\nu&N^c 
\end{array}
\right)\times
 \left(
\begin{array}{cc}
0  & m \\
m & M
\label{era4}
\end{array}
 \right)
\times
\left(
\begin{array}{c}
\nu\\
N^c 
\label{era5}
\end{array}
\right),
\label{er1245}
\eeqa
where
\beq
m= \lambda_1  \upsilon.
\label{eigen1}
\eeq
After diagonalization
the neutrino mass matrix gives us two eigenvalues,
the heavy eigenvalue
\beq
m_{heavy} \approx M =\ \lambda_2 \frac{<H>^2 }{M_s},
\label{neu2}
\eeq
corresponding to the right handed neutrino and 
the light eigenvalue
\beq
m_{light} \approx \frac{m^2}{M} =\ \frac{\lambda_1^2}{\lambda_2  }
\times\frac{\upsilon^2 \ M_s  } { <H>^2} 
\label{neu1}
\eeq
corresponding
to the left handed neutrino \footnote{
The neutrino mass matrix is of
the type of an extended Frogatt- Nielsen mechanism mixing
light with heavy states.}. In the present models, $\lambda_1 =1$.

Values of the parameters giving us values
for neutrino 
masses between 0.1-10 eV, consistent
with the observed neutrino mixing in neutrino
oscillation measurements, will not be presented here, as they
have already been discussed in \cite{kokos1}.
The analysis remain the same, as the mass scales of
the theory do not change. We note that 
the hierarchy of neutrino masses
has been investigated by examining several
different scenaria associated with a light
$\nu_L$ mass including the cases
$ \langle H \rangle = |M_s|$,
$ \langle H \rangle < | M_s |$.
In both cases the hierarchy of neutrino masses
is easily obtained.

\subsection{\em Exotic fermion couplings}

Our main focus in this part
is to show that
all additional particles, appearing in table
(\ref{spectrum8}), beyond those of SM get a
heavy mass
and disappear from the low energy spectrum.
The only exception will be the light masses of
$\chi_L^1$, $\chi_L^2$, weak fermion doublets 
which are of order of the electroweak symmetry
breaking scale, e.g. 246 GeV.

Lets us discuss the latter issue in more detail.
\newline
The left handed fermions $\chi_L^1$ receive a mass
from the coupling 
\beq
(1, 2, 1)(1, 2, 1) e^{-A}
 \frac{\langle h_2 \rangle \langle h_2 \rangle
\langle {\bar F}_R^H  \rangle \langle H_1 \rangle
\langle s_B^1 \rangle}{M_s^4}
\stackrel{A \rightarrow 0}{\sim}
\frac{\upsilon^2}{M_s} \ (1, 2, 1)(1, 2, 1)
\label{ka1sa1}
\eeq
that is in representation form
\beqa
(1, 2, 1)_{(0, 1, 0, 1, 0)} \ (1, 2, 1)_{(0, 1, 0, 1, 0)} 
\langle (1, {\bar 2}, {\bar 2})_{(0, -1, -1, 0, 0)} \rangle \
\langle(1, {\bar 2}, {\bar 2})_{(0, -1, -1, 0, 0)} \rangle &\nonumber\\
\times \ \langle({\bar 4}, 1, 2)_{(-1, 0, 1, 0, 0)}\rangle \
\langle(4, 1, 2)_{(1, 0, 1, 0, 0)}\rangle \ 
\langle (1, 1, 1)_{(0, 0, 0, -2, 0)}\rangle 
\label{ka1sa111}
\eeqa
In (\ref{ka1sa1}) we have included the leading contribution of the 
worksheet area connecting the
seven vertices. In the following for simplicity reasons we will set the 
leading contribution of the different couplings to
one e.g. area tends to zero.

The left handed fermions $\chi_L^2$
receive an order $M_s$  mass
from the coupling 
\beq
(1, 2, 1)(1, 2, 1) 
 \frac{\langle h_2 \rangle \langle h_2 \rangle
\langle {\bar F}_R^H  \rangle \langle H_1 \rangle
\langle  {\bar s}_B^2 \rangle}{M_s^4}
\stackrel{A \rightarrow 0}{\sim}
\frac{\upsilon^2}{M_s} \ (1, 2, 1)(1, 2, 1)
\label{ka1sa2}
\eeq
that is in representation form :
\beqa
(1, 2, 1)_{(0, 1, 0, 0, -1)} \ (1, 2, 1)_{(0, 1, 0, 0, -1)} 
\langle (1, {\bar 2}, {\bar 2})_{(0, -1, -1, 0, 0)} \rangle \
\langle(1, {\bar 2}, {\bar 2})_{(0, -1, -1, 0, 0)}
\rangle &\nonumber\\
\times \ \langle({\bar 4}, 1, 2)_{(-1, 0, 1, 0, 0)}
\rangle \
\langle(4, 1, 2)_{(1, 0, 1, 0, 0)}\rangle \ 
\langle (1, 1, 1)_{(0, 0, 0, 0, 2)}\rangle \
\label{ka1sa112}
\eeqa
Altogether, $\chi_L^1$, $\chi_L^2$,  
 receive of order $\upsilon^2/M_s$. Thus the Pati-Salam models
 predict light weak doublets with mass between 90 and
 $\upsilon = 246$ GeV.
 This
 is a general
 prediction of all classes of models based on intersecting
 D6-branes.

The $\chi_R^1$ doublet fermions receive heavy masses
of order $M_s$ in the following way.
The mass term
\beq
(1, 1, 2)(1, 1, 2)\frac{\langle H_2 \rangle
\langle F_R^H \rangle 
\langle s_B^1 \rangle}{M_s^2}
\label{real21}
\eeq
can be realized.
In explicit representation form
\beqa
 (1, 1, 2)_{(0, 0, 1, -1, 0)}
 \ (1, 1, 2)_{(0, 0, 1, -1, 0)} \
\langle ({\bar 4}, 1, {\bar 2})_{(-1, 0, -1, 0, 0)} \rangle \
\langle ({4}, 1, {\bar 2})_{(1, 0, -1, 0, 0)} \rangle
\nonumber\\
\times \langle (1, 1, 1)_{(0, 0, 0, 2, 0}\rangle 
\label{real200}
 \eeqa
With vevs $<H_2> \sim <F_R^H>  \sim M_s$,
the mass of $\chi_R^1$ is of order $M_s$.

The $\chi_R^2$ doublet fermions receive heavy masses
of order $M_s$ in the following way:
\beq
(1, 1, 2)(1, 1, 2)\frac{\langle H_2 \rangle
\langle F_R^H \rangle 
\langle s_B^2 \rangle}{M_s^3}
\label{real211}
\eeq
In explicit representation form
\beqa
 (1, 1, 2)_{(0, 0, 1, 0, 1)}  \ (1, 1, 2)_{(0, 0, 1, 0, 1)} \
\langle ({\bar 4}, 1, {\bar 2})_{(-1, 0, -1, 0, 0)} \rangle \
\langle (4, 1, {\bar 2})_{(1, 0, -1, 0, 0)} \rangle
\nonumber\\ \times
\langle (1, 1, 1)_{(0, 0, 0, 0, -2}\rangle  
\label{real2001}
 \eeqa
With vevs $<H_2> \sim <F_R^H>  \sim M_s$,
the mass of $\chi_R^2$ is of order $M_s$.

The 6-plet fermions, $\omega_L$, receive a mass term of
order $M_s$  from the
coupling, 
\beq
({\bar 6}, 1, 1)({\bar 6}, 1, 1)
\frac{\langle H_1  \rangle \langle {F}_R^H
\rangle \langle H_1 \rangle \langle {F}_R^H
\rangle}{M_s^3}
\label{6plet}
\eeq
where we have made use of the $SU(4)$ tensor products
$6 \otimes 6 = 1 + 15 + 20$, $ 4 \otimes 4 = 6 + 10$.
Explicitly, in representation form,
\beqa
({\bar 6}, 1, 1)_{(-2, 0, 0, 0, 0)}
\ ({\bar 6}, 1, 1)_{(-2, 0, 0, 0, 0)}
\langle(4, 1, 2)_{(1, 0, 1, 0, 0)} \rangle \
\langle(4, 1, 2)_{(1, 0, 1, 0, 0)} \rangle &\nonumber\\
\times \ \langle (4, 1, {\bar 2})_{(1, 0, -1, 0, 0)})
\rangle \ \langle(4, 1, {\bar 2})_{(1, 0, -1, 0, 0) })
\rangle
\label{6plet1}
\eeqa

The 10-plet fermions $z_R$ receive
 a heavy mass of order $M_s$ from the coupling
\beq
(10, 1, 1)(10, 1, 1)\frac{\langle {\bar F}_R^H
\rangle \langle {\bar F}_R^H \rangle \langle H_2
\rangle \langle H_2 \rangle}{M_s^3},
\label{10plet}
\eeq
where we have used the
tensor product representations for $SU(4)$,
$10 \otimes 10 = 20 + 35 + 45$,
$20 \otimes {\bar 4} = {\bar 15 } + {\bar 20}$, 
${\bar 20} \otimes {\bar 4} = {\bar 6 } + 10$,
$10 \otimes {\bar 4} =  4  + 36$, $4 \otimes {\bar 4} = 1 + 15$.
Explicitly, in representation form, 

\beqa
(10, 1, 1)_{(2, 0, 0, 0, 0)} (10, 1, 1)_{(2, 0, 0, 0, 0)}
\langle({\bar 4}, 1, 2)_{(-1, 0, 1, 0, 0)}\rangle \
\langle({\bar 4}, 1, {2})_{(-1, 0, 1, 0, 0)}\rangle
&\nonumber\\
\times  \
\langle({\bar 4}, 1, {\bar 2})_{(-1, 0, -1, 0, 0)}\rangle \
\langle({\bar 4}, 1, {\bar 2})_{(-1, 0, -1, 0, 0)}\rangle
\label{10pletagain}
\eeqa


\section{Conclusions}

Recently the first examples
of three generation string
GUT models
that break exactly to the SM at low energies were
constructed \cite{kokos1}. These models
were based
on the $SU(4)_C \times SU(2)_L \times SU(2)_R$ structure
at the string scale.
They are build on a background
of intersecting D6-branes wrapping on 1-cycles across
each of the
three $T^2$-tori appearing in the decomposition $T^6 = T^2 \times 
T^2 \times T^2$ in IIA orientifolds \cite{tessera}.  
In this work, we extended the four stack
constructions of \cite{kokos1}
to five stacks. The different classes of GUT models are
constructed
as
deformations around the basic intersection
structure of the accommodated quark and lepton
representations.
Thus the massless structure of the
quark-lepton and
the Higgs sector is being shared by the present
PS-II GUTS and also by
the Pati-Salam PS-A, PS-I \cite{kokos1} classes of
$SU(4) \times SU(2)_L \times SU(2)_R$ GUT models.

The new classes of models preserve several features of the 
original models \cite{kokos1}. 
Among them we mention that the proton is stable as the
baryon number is a gauged summetry, the
corresponding gauge boson become massive
through its BFF couplings, and thus baryon number survives
as a global symmetry to low energies.
Particularly important in the satisfaction of the RR
tadpole
cancellation conditions is the addition of extra
$U(1)$ branes. This is to be contrasted 
with models with just the SM at low energy from an
extended Standard model structure
at the string scale \cite{louis2, kokos, kokos2, D5, D51}.
In those
cases the presence of
extra branes has no intersection with the rest of the
branes \footnote{Also in this case the extra branes can
be characterized as
hidden one's as they don't charge the chiral fermion
context of the models.}.
In the present constructions the extra branes are
handled in such a way that their presence
has non-trivial intersection numbers with the colour
$a$- and the
leptonic $U(1)$ $d$-, $e$- branes. The presence of extra
branes creates scalar singlet
scalars 
that may be used to break the additional extra $U(1)$'s
that
survive massless the Green-Schwarz mechanism.   

Also in the construction
of the models
we allow exotic, antisymmetric and symmetric,
fermionic representations of the colour, and $U(1)$
degrees of
freedom arising from
brane-orientifold
image brane sectors. 
In this way we engineer the models such that they have
the capacity to accommodate couplings
that give a mass of 
order $M_s$ to all these exotic fermions, and also
create singlets that may be used to break the exra U(1)'s
surviving massless the presence of the
generalized Green-Schwarz
mechanism.

Small neutrino masses,
of order 0.1-10 eV, in
consistency with neutrino oscillation experiments,
can be easily accommodated as the worldsheet area, involved in the
Yukawa couplings,
between the intersecting branes works practically
as a moduli parameter. 
Moreover,
colour triplet Higgs couplings that could couple
to quarks and leptons
and cause a problem to proton decay are absent in all
classes of models.

The present non-supersymmetric model constructions
if the angle
stabilization conditions of Appendix I hold, are
free of tachyons. However, this is not enough as
there will always be closed string NSNS tadpoles 
that cannot all be removed at once.
Some ways that these tadpoles may be removed
have been
suggested in
\cite{sugge} by
freezing
the complex moduli
to discrete values, or by background redefinition
in terms of 
wrapped metrics \cite{nsns}. However, 
a dilaton tadpole always remain
that could in principle
reintroduce tadpoles in the next leading order.
Forcing us to rethink a solution in terms of the
Fischer-Susskind mechanism \cite{fi}.
We also note that the complex structure
moduli \footnote{ As was noted in \cite{kokos1}
the
K\"ahler moduli could be
fixed from its value at the string scale, using relations
involving the product radii (see (\cite{kokos1}) )
but in this way we could
use a large fine tuning which seems unnatural in
a string theory context,
where moduli should be assigned values dynamically.}
can be fixed
to discrete values using the supersymmetry conditions, e.g. see (\ref{condo3}),
and in this way it is possible that some if not all, of the NS tadpoles 
can be removed. We leave this task for a future investigation.
Also related is the fact that we have tacitly assumed that
the values of the scalar singlets present in the models
are of the order of the string scale.
In principle, whether these scalars, appearing in $N=1$
supermultiplets, really receive a vev is highly non-trivial dynamical
question. A full solution to the problem
involves a) a solution to the stability of the present
non-SUSY backgrounds and also b) a determination by a string
calculation of the full effective potential of the scalar
fields. Both problems are open problems as involve non-trivial
dynamics and are beyond our calculational ability at
present.

One point that we want to emphasize is that until
recently, in
orientifolded
six-torus compactifications   
there was not any obvious explanation
for keeping the string scale low \cite{antoba}, e.g. to
the 1-100 TeV
region.
Thus the usual explanation of explaining the
hierarchy 
by making the Planck scale large, while keeping the
string scale low,
 by varying the radii of the 
transverse directions \cite{antoba} could not be
applied \footnote{
As there are no
simultaneously transverse torus directions to 
all D6-branes \cite{tessera}.} .
However, as was noted in \cite{kokos1}
there is an alternative mechanism
that keeps the string scale $M_s$ low. In particular the existence
of the light weak doublets $\chi_L^1$, $\chi_L^2$  with mass of order 
$\upsilon^2/M_S$ and up to
246 GeV, makes a definite prediction for a low string scale 
in the energy range less than 650 GeV if a fermion weak doublet has to be 
detected in $e^{+}e^{-}$ experiments, in the energy range over 90 GeV.
If the lightest weak fermion doublet is over 100 GeV as being favoured by 
the experiments at present the $M_s < 600 GeV$.
That effectively, makes the PS-II class of D6-brane 
models (also the PS-A, PS-I classes) directly 
testable to present or feature accelerators. \newline   
Also it would be interesting if we could analyze the low energy 
implications for the PS-I, PS-A, PS-II GUT models in terms of a variant of
 the 
analysis performed in \cite{que}.

\begin{center}
{\bf Acknowledgments}
\end{center}
I am grateful
to Luis Ib\'a\~nez,
and Angel Uranga,
for useful discussions. 

\newpage

\section{Appendix I}

In the appendix we list the conditions, mentioned in
section 5,
  under which the PS-II model D6-brane configurations
of tadpole solutions of table (\ref{spectruma101}),
are tachyon free.
Note that 
the conditions are expressed in terms of the angles
defined in (\ref{angPSII}).
A comment is in order.
Note that we have included the contributions
from the sectors $ab^{\star}$, $ac$, $bd^{\star}$,
$cd$, $be$, $ce^{\star}$. We have not present the
tachyon free conditions from the sectors $dd^{\star}$,
$ee^{\star}$, as these conditions will be shown to be
equivalent to the
presence of N=1 supersymmetry in these sectors.

\beqa
\begin{array}{ccccccc}
-(\frac{\pi}{2} + \vartheta_1) &+& \vartheta_2 &+& 2 {\vartheta}_3 &\geq& 0\\
-(\frac{\pi}{2}-{\tilde \vartheta}_1) &+& \vartheta_2 &+& 
2{\vartheta}_3 &\geq &0 \\
-(\frac{\pi}{2}+\vartheta_1) &+& {\tilde \vartheta}_2
&+& 2 {\vartheta}_3 &\geq& 0\\
-(\frac{\pi}{2} - {\tilde \vartheta}_1) &
+&{\tilde \vartheta}_2 &+&
2{\vartheta}_3 &\geq& 0 \\
-(-\frac{\pi}{2} + { \vartheta}_1) &+&{\bar
\vartheta}_2 &+&
2 {\vartheta}_3 &\geq& 0 \\
-(\frac{\pi}{2} + {\tilde \vartheta}_1) &+&{\bar
\vartheta}_2 &+&
2{\vartheta}_3 &\geq& 0 \\
\\\\
(\frac{\pi}{2} + \vartheta_1) &-& \vartheta_2 &+& 2
{\vartheta}_3 &\geq& 0\\
(\frac{\pi}{2}-{\tilde \vartheta}_1) &-& \vartheta_2 &+& 
2{\vartheta}_3 &\geq &0 \\
(\frac{\pi}{2}+\vartheta_1) &-& {\tilde \vartheta}_2
&+& 2 {\vartheta}_3 &\geq& 0\\
(\frac{\pi}{2} - {\tilde \vartheta}_1) &
-&{\tilde \vartheta}_2 &+&
2{\vartheta}_3 &\geq& 0 \\
(-\frac{\pi}{2} + { \vartheta}_1) &-&{\bar
\vartheta}_2 &+&
2 {\vartheta}_3 &\geq& 0 \\
(\frac{\pi}{2} + {\tilde \vartheta}_1) &-&{\bar
\vartheta}_2 &+&
2{\vartheta}_3 &\geq& 0
\end{array}
\eeqa

\beqa
\begin{array}{ccccccc}
(\frac{\pi}{2} + \vartheta_1) &+& \vartheta_2
&-& 2 {\vartheta}_3 &\geq& 0\\
(\frac{\pi}{2}-{\tilde \vartheta}_1) &+& \vartheta_2 &-& 
2{\vartheta}_3 &\geq &0 \\
(\frac{\pi}{2}+\vartheta_1) &+& {\tilde \vartheta}_2
&-& 2 {\vartheta}_3 &\geq& 0\\
(\frac{\pi}{2} - {\tilde \vartheta}_1) &
+&{\tilde \vartheta}_2 &-&
2{\vartheta}_3 &\geq& 0 \\
(-\frac{\pi}{2} + { \vartheta}_1) &+&{\bar
\vartheta}_2 &-&
2 {\vartheta}_3 &\geq& 0 \\
(\frac{\pi}{2} + {\tilde \vartheta}_1) &+&{\bar
\vartheta}_2 &-&
2{\vartheta}_3 &\geq& 0 \\
\label{free}
\end{array}
\eeqa

\newpage

\section{Appendix II}
In this Appendix,
following a comment in (\ref{loc1})
we emphasize the importance of choosing
an appropriate location for the presence of
extra branes needed
to satisfy the RR tadpole cancellation conditions.
By choosing the location of the extra branes, in a
general point
like $(1/\beta_1, 0)(1/\beta_1, 0)(1, m/2)$ and
e.g. $\beta_1 = \beta_2 = 1/2$, $m=1$,
we are getting a GUT class of models where there are no
electroweak bidoublets $h_1, h_2, h_3, h_4$ and thus
there is no Dirac term
allowed to give mass to quarks and leptons.

The precise arguments have as follows :
In this case the number of extra branes required is
$N_h= 4$.
The structure of U(1) anomalies gives us the
following couplings of the RR fields to the U(1)'s of the
new classes of models :

\beqa
B_2^3 \wedge 
[2 {\tilde \epsilon}]
[-(F^b +\ F^c) + (F^{h_1} +\ F^{h_2} +\ F^{h_3} +\ F^{h_4})],&\nonumber\\
B_2^1 \wedge [\epsilon] 
[ 4 n_a^2 \ F^a +
4 m_b^1 \ F^b + 4 m_c^1 \ F^c + 2 n_d^2  F^d +
2 n_e^2 F^e],&\nonumber\\
B_2^o  \wedge \left(   3  F^a - 2 F^d + F^e  \right) .&
\label{newPSIIb}
\eeqa

The couplings of the dual scalars $C^I$ of $B_2^I$,
required to cancel
the mixed anomalies of the $U(1)$'s with the 
non-abelian gauge groups $SU(N_a)$,
are given by
\beqa
C^o \wedge 2  [-(F^b \wedge F^b)
+ (F^c \wedge F^c) + 2(F^{h_1} \wedge F^{h_1} +
F^{h_2} \wedge F^{h_2} + F^{h_3} \wedge F^{h_3} +
F^{h_4} \wedge F^{h_4} )], &\nonumber\\
C^2 \wedge [{\epsilon}{\tilde \epsilon}][
  2 n_a^2 (F^a \wedge
F^a)  +  2 m_b^1 (F^b \wedge 
F^b) -  2 m_c^1 (F^c \wedge F^c) 
+ n_d^2 (F^d \wedge F^d) &\nonumber\\
- n_e^2 (F^e
\wedge F^e)],& \nonumber\\
C^0 \wedge [{\epsilon}{\tilde \epsilon}][
  2 n_a^2 (F^a \wedge F^a) -4
  F^d \wedge F^d) -4 (F^e \wedge F^e) ],& \nonumber\\
\label{newdual}
\eeqa

As can be seen two anomalous
combinations of $U(1)$'s, e.g.
 $3 F^a - 2 F + F^e$, $-(F^b + F^c) + F^{h_1} + F^{h_2} + F^{h_3} +
 F^{h_4}$
 become massive through their couplings to RR
 fields $B_2^o$, $B_2^3$. Also there is an anomaly free
 model dependent U(1) which is getting massive from
 its coupling to the RR field $B_2^1$. 
In addition, there are six non-anomalous $U(1)$'s
which also are
getting broken by vevs of singlet scalars generated
by imposing N=1 SUSY on certain sectors.
They are :
\beqa
U(1)^{(4)} =\ {\tilde \epsilon}(F^{h_1} -\ F^{h_2} +\ F^{h_3} -\ F^{h_4}),&\nonumber\\
U(1)^{(5)} =\ {\tilde \epsilon}(F^{h_1} -\ F^{h_2} -\ F^{h_3} +\ F^{h_4})                           &\nonumber\\
U(1)^{(6)} =\ (Q_b - Q_c) + (Q_a + Q_d - Q_e) +
{\tilde \epsilon}(F^{h_1} +\ F^{h_2} -\ F^{h_3} -\ F^{h_4})&\nonumber\\
U(1)^{(7)} =\ \frac{1}{5}[(Q_b - Q_c) + (Q_a + Q_d -Q_e)]
+ \frac{1}{4} {\tilde \epsilon}(-F^{h_1} -\ F^{h_2}
+\ F^{h_3} +\ F^{h_4})            &\nonumber\\
U(1)^{(8)} =\ (Q_b + Q_c) + \frac{1}{2} (F^{h_1} +\ F^{h_2}
+\ F^{h_3}
+\ F^{h_4})         &\nonumber\\
U(1)^{(9)} =\ Q_a + 4 Q_d + 5 Q_e       &
\label{tora1}
\eeqa
The choice of U(1)'s (\ref{tora1})
gives the constraints 
\beq
m_b^1 = - m_c^1
\label{hola1}
\eeq
\beq
2 n_a^2 +\ 4 n_d^2 +\ 5 n_e^2 =\ 0
\label{hola2}
\eeq
The number of electroweak bidoublets $h_1$, $h_2$
taking into
account the constraint (\ref{hola2})
is zero. We note that this numbers depends on the
difference $|m_b^1 -m_c^1|$. 
Also the number of electroweak bidoublets $h_3$, $h_4$ is zero,
 as it
depends on the difference $|m_b^1 + m_c^1|$.

\end{document}